\documentclass[10pt]{article}
\usepackage[explicit]{titlesec}
\usepackage{hyphenat}
\usepackage{ragged2e}
\RaggedRight

\usepackage{rotating}  
\usepackage{theorem}   
\usepackage{bm}  
\usepackage{amsmath,amsfonts,amssymb}
\usepackage{booktabs}  
\usepackage{pdfsync}   
\usepackage{graphicx}
\usepackage{latexsym}
\usepackage{amsmath}
\usepackage{amssymb}
\usepackage{mathtools}

\usepackage{fancyvrb}
\usepackage{subfigure}
\usepackage{colortbl}
\usepackage{enumerate}
\usepackage{pifont}
\usepackage{stmaryrd}
\usepackage{textcomp}
\usepackage{fncylab}
\usepackage{multirow}
\usepackage{paralist}
\usepackage{wrapfig}
\usepackage{longtable}
\usepackage[]{algorithm2e}
\usepackage[table]{xcolor}
\usepackage{lscape}
\usepackage{lipsum}
\usepackage{amsmath}
\usepackage{csquotes}

\usepackage{garamondx}
\usepackage[T1]{fontenc}
\usepackage{amsmath}
\usepackage{graphicx}
\usepackage{xcolor}

\usepackage{geometry}
\usepackage{fancyhdr}
\usepackage[labelfont={footnotesize,bf} , textfont=footnotesize]{caption}
\makeatletter
\renewcommand\tagform@[1]{\maketag@@@ {\ignorespaces {\footnotesize{\textbf{Equation}}} #1.\unskip \@@italiccorr }}
\makeatother
\titleformat{\section}
  {\normalfont}{\thesection}{1em}{\MakeUppercase{\textbf{#1}}}
\titlespacing\section{0pt}{0pt}{-10pt}
\titleformat{\subsection}
  {\normalfont}{\thesubsection}{1em}{\textit{#1}}
\titlespacing\subsection{0pt}{0pt}{-8pt}

\makeatletter
\newcommand\sixteen{\@setfontsize\sixteen{17pt}{6}}
\renewcommand{\maketitle}{\bgroup\setlength{\parindent}{0pt}
\begin{flushleft}
\sixteen\bfseries \@title
\medskip
\end{flushleft}
\textit{\@author}
\egroup}
\makeatother

\usepackage[sort&compress]{natbib}
\makeatletter
\renewcommand\@biblabel[1]{\textbf{#1.}\hfill}
\makeatother

\bibpunct{}{}{,~}{s}{,}{,}
\title{Towards Mining Creative Thinking Patterns from Educational Data}

\author{
Nasrin Shabani*$^{a}$\\ \medskip
$^{a}$Macquarie University, Sydney, Australia\\  \medskip
nasrin.shabani@mq.edu.au
}

\begin{document}

\vspace*{.01 in}
\maketitle
\vspace{.12 in}

\section*{abstract}

Creativity, i.e., the process of generating and developing fresh and original ideas or products that are useful or effective, is a valuable skill in a variety of domains. Creativity is called an essential 21st-century skill that should be taught in schools. The use of educational technology to promote creativity is an active study field, as evidenced by several studies linking creativity in the classroom to beneficial learning outcomes. Despite the burgeoning body of research on adaptive technology for education, mining creative thinking patterns from educational data remains a challenging task. In this paper, to address this challenge, we put the first step towards formalizing educational knowledge by constructing a domain-specific Knowledge Base to identify essential concepts, facts, and assumptions in identifying creative patterns. We then introduce a pipeline to contextualize the raw educational data, such as assessments and class activities. Finally, we present a rule-based approach to learning from the Knowledge Base, and facilitate mining creative thinking patterns from contextualized data and knowledge. We evaluate our approach with real-world datasets and highlight how the proposed pipeline can help instructors understand creative thinking patterns from students' activities and assessment tasks.

\section*{keywords}
Educational Data Mining, Creativity Patterns, Data Curation

\vspace{.12 in}


\section{introduction}

In this section, we first begin with an overview of the work and explain the problem statement which is understanding and identifying creative thinking patterns from raw educational data. Next, we provide our contributions and explain how the proposed method greatly facilitates the process of discovering creativity patterns. Finally, we outline the organization of this paper.

\subsection{Overview and Problem Statement}
Around the world today, original and creative ideas are the best and most important products of any powerful country. This clearly shows the importance of recognizing and nurturing creativity in children from a young age. Creativity is a set of skills that all humans have the capacity to possess, but it must be nurtured and expanded under the right circumstances. Unlike earlier theories that assumed creativity as an inherited and intrinsic process, recent research on creativity via education reveals that creative thinking, i.e., the ability to consider something in a new way, is considered a skill and can be learned by individuals.
Therefore, in countries with a dynamic education system, fostering creativity has been considered one of the most important goals of education~\cite{shaheen2010creativity}.

Creativity is defined in many different ways, but almost all researchers and psychologists consent to a definition that describes it as a process of generating ideas or products that are both unique and useful~\cite{amabile2018creativity}.
It may involve
a set of challenges, such as generating various solutions in response to a problem, spotting problems in the existing state of knowledge, dealing with ambiguous circumstances, and bringing ideas into action. Hence, in the process of creative pursuits, students need ongoing guidance, and training. But before providing personalized training, it is needed to first detect creativity patterns in students~\cite{muldner2015utilizing}.

There are numerous
existing tools and techniques
for measuring and detecting creativity such as divergent thinking tests~\cite{torrance1972predictive}, self-report measures of creativity~\cite{miller2014self}, or judgment of products~\cite{amabile2018creativity} which have been traditionally used to identify creative students. These approaches mostly involve evaluating the quality of ideas and products. Practically,
using human evaluators
to asses students' responses to creative tasks, such as rating the uniqueness of ideas from the alternate uses test, is a common element of doing creativity research. Although scoring systems have proven effective, they are susceptible to two fundamental limitations: labor cost and subjectivity, which pose specific psychometric risks to reliability and validity~\cite{beaty2021automating}.

Recent development in learning management systems has proved to assist different educational assessments and minimize the aforementioned limitations. These technologies have the capacity to gather and visualize large amount of educational data,  e.g., assessments and class activities.
In this context, helping instructors to understand the educational data
remains a challenging task~\cite{diana2017instructor}.
There is a large amount of work aiming to
discover insight from educational data, with the goal to support traditional learning and educational assessments~\cite{ baker2010datamining,romero2014survey,romero2020educational, bull2004open}.
However, the current state of the art lacks a significant data-driven strategy to link students' behaviour to creativity patterns.

In this paper, to address these challenges, we  construct a domain-specific Knowledge Base (KB) to identify essential concepts, facts, and assumptions in identifying creative patterns. We then introduce a pipeline to contextualize the raw educational data, such as assessments and class activities. Finally, we present a rule-based approach to learning from the Knowledge Base, and facilitate mining creative thinking patterns from contextualized data and knowledge. We evaluate our approach with real-world datasets and highlight how the proposed framework can help instructors understand creative thinking patterns from students' activities and assessment tasks.

\subsection{Contributions}
In this study, we propose a rule-based insight discovery method to discover patterns of creativity in educational data. Our work relies on the knowledge of experts in education for buildings a domain specific Knowledge Base
to be linked to the extracted features from educational data. To achieve this goal, we first imitate the knowledge of domain experts into an Educational Knowledge Base, i.e., a set of concepts organized into a taxonomy, instances for each concept, and relationships among them.
Secondly, to drive insight from raw data we propose a method
to link the concept nodes in the taxonomy to the entities extracted from
educational data.

Our approach is based on a motivating scenario in educational assessment, where a knowledge worker (e.g., a teacher) may need to analyze the activities
of students in a classroom and augment that information with the knowledge in the educational Knowledge Base.
Making benefit from a user-guided rule-based technique the person can link the information extracted from raw educational data to creativity patterns, identified in the educational Knowledge Base.

The unique contributions of this paper are:

\begin{itemize}
    \item We put the first step towards formalizing the educational knowledge by constructing a domain-specific Knowledge Base to identify essential concepts, facts, and assumptions in identifying creative patterns.
    \item We introduce a pipeline to contextualize the raw educational data, such as assessments and class activities. We customize existing data curation techniques to turn the raw educational data into contextualized data and knowledge.
    \item we present a rule-based approach to learn from the Knowledge Base, and facilitate mining creative thinking patterns from contextualized data and knowledge.
    \item We evaluate our approach with real-world datasets and highlight how the proposed framework can help instructors understand creative thinking patterns from students' activities and assessment tasks.
\end{itemize}

\subsection{Summary and Outline}

In this section, we gave a broad picture of the problem and covered the motivation, problem statement, and contributions of this paper.
The remaining of this paper is structured in the following manner:

\begin{itemize}

\item
In Section~2 we will present the background and analyze the
related work in the education domain. We will review the
current state-of-the-art approaches, including educational knowledge, educational modelling, Open Learner Models, and Educational Data Mining techniques. We will provide a summary to compare the reviewed methods and techniques in the four areas.

\item
In Section~3 we will explain our methodology towards mining creative thinking patterns from educational data. This section will provide detailed information about the proposed method which includes building an Educational Knowledge Base, data curation, feature selection, building a Knowledge Graph, and finally linking the graph to the proposed knowledge base using a novel rule-based technique.

\item
In Section~4 we will present the experiment, results, and evaluation. We will discuss a motivating scenario to clarify our approach to the reader and then explain the experiment's dataset and setup. We will conclude this section with a discussion over the evaluation results.

\item
Finally, in section~5, we will conclude the study by providing a summary of the proposed method. We will also discuss the future directions that we build on our study.

\end{itemize}

\section{Background and State-of-the-Art}
In this section, we study and analyze the recent work in modelling and mining educational data.
We introduce the key terms and background in the field, and then discuss the related works in Educational Data, Educational Data Modeling, Open Learner Models, Educational Data Mining, and Learning Analytics. We conclude the section by summarizing the challenges in educational data mining and analytics, and highlight the added value of our proposed approach.

\subsection{Data in Education}

\subsubsection{Data, Metadata, and Big Data}

\textbf{Data} in computing refers to the information that
is converted to a digital format that could be
stored and processed. Individuals, machines, and sensors,
as sources of information, are generating a massive amount of data each second. For example, individuals generate data when they take a photo using their digital devices, browse the internet, or publish content on their social media. Smart devices such as cars, watches, TVs, and phones also sense and create data even when they are not in use~\cite{beheshti2020intelligent,schiliro2018icop}.

Data can be stored in three formats: structured, unstructured, and semi-structured. Structured data consists of data types that are well defined and have patterns that make them easy to search including student name, ID number, and age. Unstructured data consists of data types that are often difficult to search, including text, pdf, image, and video. And, semi-structured data consists of a loosely organized meta-level structure that can contain unstructured data, e.g.,
Email, HTML, XML, and JSON documents. From the processing point of view, data processing consists of organizing, curating, analyzing, and presenting operations~\cite{beheshti2018corekg}.

\textbf{Metadata} refers to as the information
that describes other data. It is also defined as a prefix that provides an 'underlying definition or description' in most information technology applications~\cite{ warren2015zen}. Metadata describes fundamental data information, making it simpler to discover, utilize, and reuse specific data instances. For example, a document file's
metadata includes the creator's information, the date that has been created or updated, and the file format.
In addition, metadata can be used for relational databases, images, videos, audio files, and web pages. By using the tracking method, smart devices such as phones or watches track our location, speed, the Apps and gadgets we are using, and the music we often listen to. Or, smart TVs can track the channels we watch, time and duration, and the Apps we use during a specific time of day~\cite{beheshti2012temporal}. Metadata offers a primary description of an information item, and it should be appropriately specified in terms of granularity, structure, quality, and provenance~\cite{alemu2015emergent,benatallah2016process,beheshti2016scalable}.

\textbf{Big Data} refers to
a set of large datasets
that are so vast, fast, or
complicated. Processing such data
using standard methods
could be difficult or impossible.
The term also refers to the technology that is used to manage the generated data and metadata. The everyday exchanged data on the
Internet,
social media contents and feeds, and mobile phone location data are examples of big data. Construction, healthcare, insurance, telecommunications, and ecommerce are among the areas where big data is becoming increasingly important. 

Wide data dispersion, a variety of formats, non-standard data models, and heterogeneous semantics are all characteristics of big data. It should be notes that several challenges are involved while using big data such as the way it should be organized, curated, analyzed, and visualized~\cite{beheshti2020intelligent}.

Big data is characterized by the following attributes (also known as four Vs~\cite{ zikopoulos2012understanding}): Volume (size of data), Velocity (speed of data streaming), Variety (data types), and Veracity (uncertainty of data).

\begin{itemize}
    \item \textbf{Volume}: The volume of data is the defining attribute of big data. Individuals, machines, and organizations create data from numerous sources, from which reliable information is collected and integrated at the organization center. For example, consider the wide range of data collected from students using their smart phones, evaluations, homework, feedback forms, surveys,
    etc. This
    mass of data creates a huge amount of data that must be appropriately analyzed.
    \item \textbf{Velocity}: This attribute of big data gives an indication of data speed, or the rate at which data is emerging from diverse sources. This high speed data is enormous and arrives in a constant stream, necessitating fast processing.
    \item \textbf{Variety}: The data that is received for processing might take many different file forms from structured to unstructured data. Formats 
    such as `.XLS' or `.XLSX' (Excel file), `.CSV'
    (comma-separated values file),
    `.TXT' (plain text file), `.PDF' (PDF file), or `.RTF'
    (rich text file). Data can arrive in a variety of formats, including audio, image, video, SMS, maps, geographical data or something we hadn't considered. It is critical for organizations to effectively handle such a diverse set of data, as there is currently a broad selection of data formats from which to extract information.
    \item \textbf{Veracity}: There are several data stream sources accessible, each of which creates a significant amount of data. Due to the large number of sources accessible, this data is susceptible to outliers or noise. As a result, the data's nature or behavior may change. Veracity deals with the uncertainty of data, which has a significant influence on the organization's decision-making process.
\end{itemize}

Recently,
the four Vs concept has been expanded into numerous Vs. For example, big data is classified into five Vs (Volume, Velocity, Veracity, and Value) in one research~\cite{ demchenko2013addressing}, while characterized into seven Vs in another one by adding two other new concepts such as Valence, and Variability~\cite{saggi2018survey}.

\subsubsection{Educational Data}

A wide range of educational data is accessible from a number of different sources. By using the educational data teachers may now monitor their students' academic achievements, learning behaviors, and offer immediate feedback based on the needs and requirements of students. Learning management systems collect a huge quantity of data from students that may be used to improve the learning environment, assist teacher in teaching and students in learning, and enhance the learning experience in general. Different learning resources are available which a number of them can be listed as follows~\cite{baradwaj2012mining}:

\begin{itemize}
    \item Interaction between students, instructors, and also students and instructors (e.g., chat boxes, discussion forums, navigation behavior).
    \item Administrative data (e.g., institution, courses, instructors),
    \item Demographic data (e.g., age, nationality, gender),
    \item Students' activities (e.g., assessments, questions, feedbacks),
    \item Students' dispositions and affectivity (e.g., attitude and motivation).
\end{itemize}

Since traditional learning analytics
are not
equipped to handle this volume of data,
big data technologies
and tools have found their way into education to process this massive amount of data. To deal with different kinds of educational
challenges
administrative data
could be
very useful. As a result, experts should acknowledge the influence of big data in education in order to alleviate educational difficulties.

In the big educational data, several research and review studies have been conducted.
Highly cited studies
investigated
themes~\cite{baig_shuib_yadegaridehkordi_2020} such as:
(i)~Behaviour and performance of learners: Dedicated to learner perspectives, fulfillment, methods, and behaviour, as well as big data structures, adaptive learning, teaching, data mining, and learning analytics~\cite{CANTABELLA2019262, chaurasia2018big};
(ii)~Educational data modeling and warehouse: Introducing big data modeling, educational data warehouses, and cloud environment research, as well as cluster analysis for educational purposes~\cite{petrova2017bigdata, ZHANG2015212};
(iii)~Improving  educational system: Introducing statistical tools, metrics, obstacles, and the usefulness of ICT. It places a strong emphasis on training and its numerous ramifications. It also establishes a rating system that monitors how websites are used in order to enhance the educational system~\cite{ong2015big, Martinez2018bigdata}; and
(iv)~Merging pedagogy with big data: Incorporating big data concepts into various courses and emphasizing their educational implications~\cite{dinter2017teaching, buffum2014cs}.

\subsubsection{Educational Environments}

The environment in which a student gets educational services is referred to as the educational environment.
There are three types of educational environments available for different users such as face-to-face learning, E-learning, and blended
learning~\cite{watson2008blend}.

\begin{itemize}
    \item \textbf{Face-to-face learning} refers to educational methods in which instructors and students meet in a certain location at a specific time for one-by-one or group class sessions, similar to what occurs in schools.
    \item \textbf{E-learning} involves any and all information delivery using digital technology for teaching and learning in education. E-learning became the foundation of modern education due to the progress of technology in information and communication. Individuals can benefit from the environment without the time and space constraints of face-to-face learning using the learning network model. Online learning, web-based learning, and virtual learning are three other terms that have also been used in literature to characterize this kind of teaching and learning environment ~\cite{goyal2012learning}.

    \item \textbf{Blended learning} practically refers to a hybrid of face-to-face and online learning that aids classroom learning.~\cite{watson2008blend}. Blended learning is viewed as a pedagogical method that integrates the effectiveness and socialization. possibilities of the classrooms with technological enhancements of E-learning.
\end{itemize}

Collecting and combining all of this raw data is a time-consuming and tough process in and of itself, therefore a preparation phase is required before modeling or applying machine learning techniques.

\subsubsection{Data Curation}

The practise of organising and integrating data from multiple sources in order to make it more useful for data analysis and discovery is known as data curation. In other words, it is the process of transforming raw data to
contextualized data and knowledge.
Data curation has been also defined as ``the active and ongoing management of data through its lifecycle of interest and usefulness''~\cite{Freitas2016}. It encompasses all of the procedures required for regulated data generation, preservation, and administration, as well as the ability to add value to data~\cite{miller2014big}. Hence, data curation should involve~\cite{beheshti2020intelligent}:

\begin{itemize}
    \item \textbf{Identifying relevant data sources}: As new data resources getting released on a daily basis, selecting the right data sources for a given analytical purpose has never been more crucial.
    \item \textbf{Ingesting data and knowledge}: The procedure for gathering information from multiple sources and transferring it to be stored in another location (e.g., database, data warehouse, or data mart) for further analysis is defined as data ingestion. There are generally two types of data ingestion: Batch data ingestion (data transfering in batches), and streaming data ingestion (collecting in real-time and transferring immediately).
    \item \textbf{Cleaning}: The process of correcting or deleting data in a database that is inaccurate, incomplete, poorly structured, or duplicated is known as data cleaning or cleansing.
    \item \textbf{Integration}: The process of combining data from many sources is called as data integration. Schema integration, identifying and addressing data value conflicts, and removing data duplication are all issues that arise during data integration.
    \item \textbf{Transformation}: Data transformation means transforming the data in such a way that is suitable for data analysis. Normalization, summarization or aggregation, smoothing, and generalization are examples of data transformation techniques~\cite{beheshti2020istory}. For instance, normalization scales the attributes within a specified range, commonly between -1.0 and 1.0, or 0.0 and 1.0. This technique is useful when the large range of attributes outweighing smaller ranges.
    \item \textbf{Adding Value}: An extremely essential step in the process of data curation that include: extraction, enrichment, annotation, linking, and summarization operations.
\end{itemize}

As data curation is one of the most significant challenges in data analytics, many organizations and researchers, first investigate converting their raw data into useful information and knowledge. Data in a variety of sources with different formats (structured to unstructured) are first stored in a storage repository called Data Lake which allows the data analysts to decide afterward on the data curation ~\cite{beheshti2019datasynapse}.

\begin{figure}[t]
	\begin{center}
	\includegraphics[width=1\linewidth]{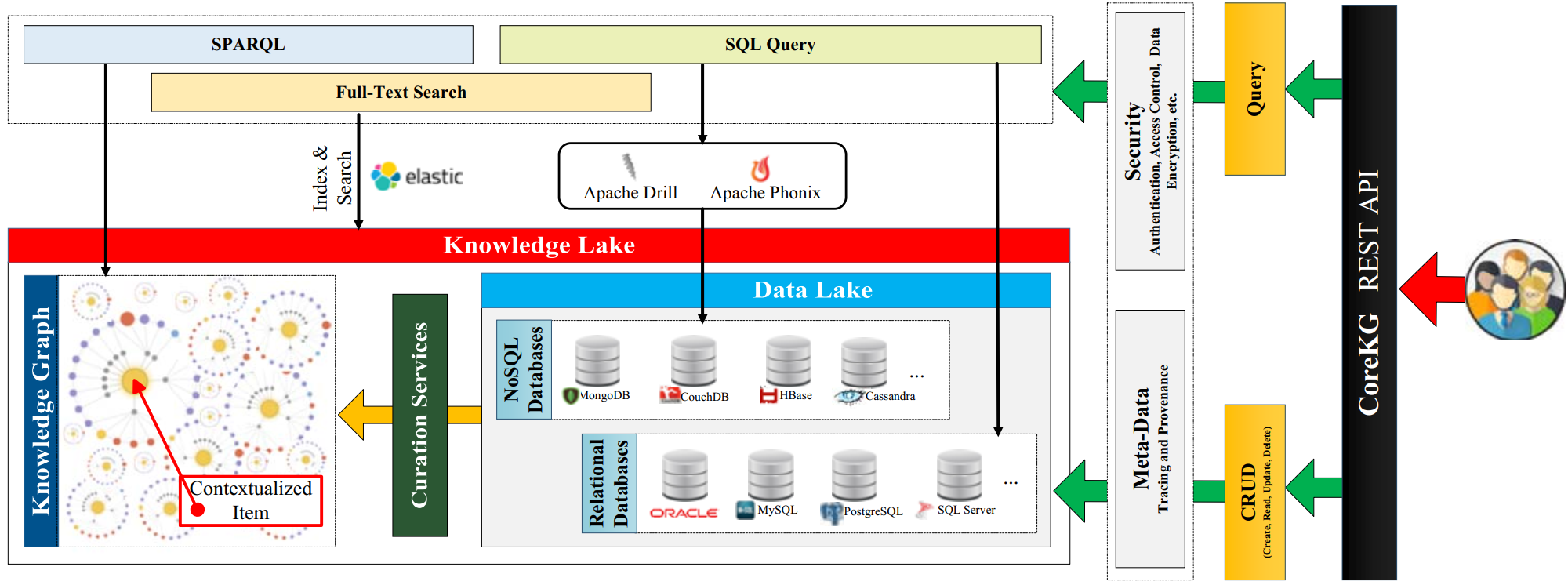}
	\end{center}
	\caption{The architecture of CoreKG~\cite{beheshti2018corekg}.}
	\label{coreKG}
\end{figure}

Many studies are now focused on curating the stored data in the Data Lakes and making this transformation intelligent, e.g, AsterixDB\footnote{http://asterixdb.apache.org/},
Orchestrate\footnote{http://orchestrate.io/}, and
CoreKG\footnote{http://github.com/unsw-cse-soc/CoreKG}~\cite{beheshti2018corekg}.
For example, recently Beheshti et al.~\cite{beheshti2018corekg,beheshti2017coredb,yang2021design} presented the Knowledge Lake or Contextualized Data Lake, which automates the curation and preparation of raw data in the Data Lake for analysis, laying the framework for big data analytics. They created CoreKG, an open source Knowledge Lake service that allows researchers and developers to manage, curate, filter, and query the data and metadata in the Lake and across time via a single REST API. The architecture and main components of the CoreKG are depicted in Figure~\ref{coreKG}.

CoreKG is a data-driven Web application development platform that powers numerous relational and NoSQL databases as a service. It enables creating islands of data (relational or/and non-relational) in the Data Lake, applying CRUD (Create, Read, Update, and Delete) on entities in those datasets, and use integrated search on top of multiple data islands. CoreKG also has a built-in architecture that protects against the most common database security vulnerabilities (e.g., Data Encryption, Access Control, and Authentication), as well as Provinence support and Tracing. In this research, they designed automated services that perform data curation and preparation in the Data Lake. Extraction, summarization, enrichment, linkage, and categorization are some of the services available.

\subsection{Educational Knowledge }

Skills and information acquired via education, practice, or experience are referred to as knowledge~\cite{brown2018knowledge}. It is also defined as recognizing a concept in a theoretical or practical manner~\cite{schleicher1999measuring}.

The educational difficulties of the 19th century highlighted the necessity for a learning knowledge foundation to make changes in the educational systems. It should not include a philosophical viewpoint, such as behaviorism or pragmatism. Instead, it should represent theories that explain the impacts on educational learning, research findings condensed into empirical results, and professional opinions about impacts on educational learning.

A variety of factors would need to be assessed such as student skills, inclinations, and past accomplishments; teacher personality and attitude in classroom; methods of teaching and resources; the time spent for teaching and learning; classroom environment; structure of the curriculum; home, school and social environment; demographic details of students; and educational policy in states and districts~\cite{wang1993toward,beheshti2020personality2vec}.

The education hierarchy refers to the hierarchical framework that most of the country's education system follows, in which students are educated with a comparable set of information based on their age group and maturity. It is involved with formalized learning and teaching in an educational system. For instance, the Australian Standard Classification of Education (ASCED) is an educational hierarchy that consists of two components, fields of education and level of education~\cite{ASED2001hierarchy}. It offers a foundation for similar administration and analytical data on educational programs and achievements grouped by level and field.

The level category consists of pre-primary education, primary education, secondary education, certificate level, diploma and advanced diploma level, bachelor degree level, graduate diploma/certificate level, and postgraduate degree level. The field category includes fields such as natural and physical science, information technology, engineering, agriculture and environmental science, health, education, etc. 

There is also another concept which is highlighted in literature, taxonomy of learning outcome. It is the process of categorizing educational activities based on the kind of related learning outcomes. A taxonomy is a categorization system that is organized in a hierarchical order which is useful to categorize phenomenons from a specific classification standpoint. Educators and educational designers utilize taxonomies to categorize the instructional objectives as well as important evaluation factors to see if the objectives are being reached or not~\cite{jonassen1998task}.

A wide variety of learning taxonomies have been designed to categorize educational statements~\cite{bloom1956taxonomy,bloom1964taxonomy,merrill1983component,gagne2005principles}. Bloom's and Bloom's revised taxonomies are two well-known taxonomies in education which assist educators evaluate students' abilities.

\textbf{Bloom's Taxonomy.}
The Bloom's taxonomy is a familiar taxonomy designed to encourage thinking, analyzing, and assessing concepts in education instead of just remembering them. It is commonly utilized in the development of teaching and learning procedures in three domains: cognitive, affective, and psychomotor~\cite{bloom1956taxonomy}. Each domain are categorized into several categories which are arranged in a progressive hierarchy from basic to complicated, with the premise that knowledge of one category leads to mastery of the next one (See Table \ref{bloom}).

\begin{table}[t]
\centering
    \begin{tabular}{l | l | l}
    Cognitive & Affective & Psychomotor \\
    \hline
    \includegraphics[width=0.3\linewidth]{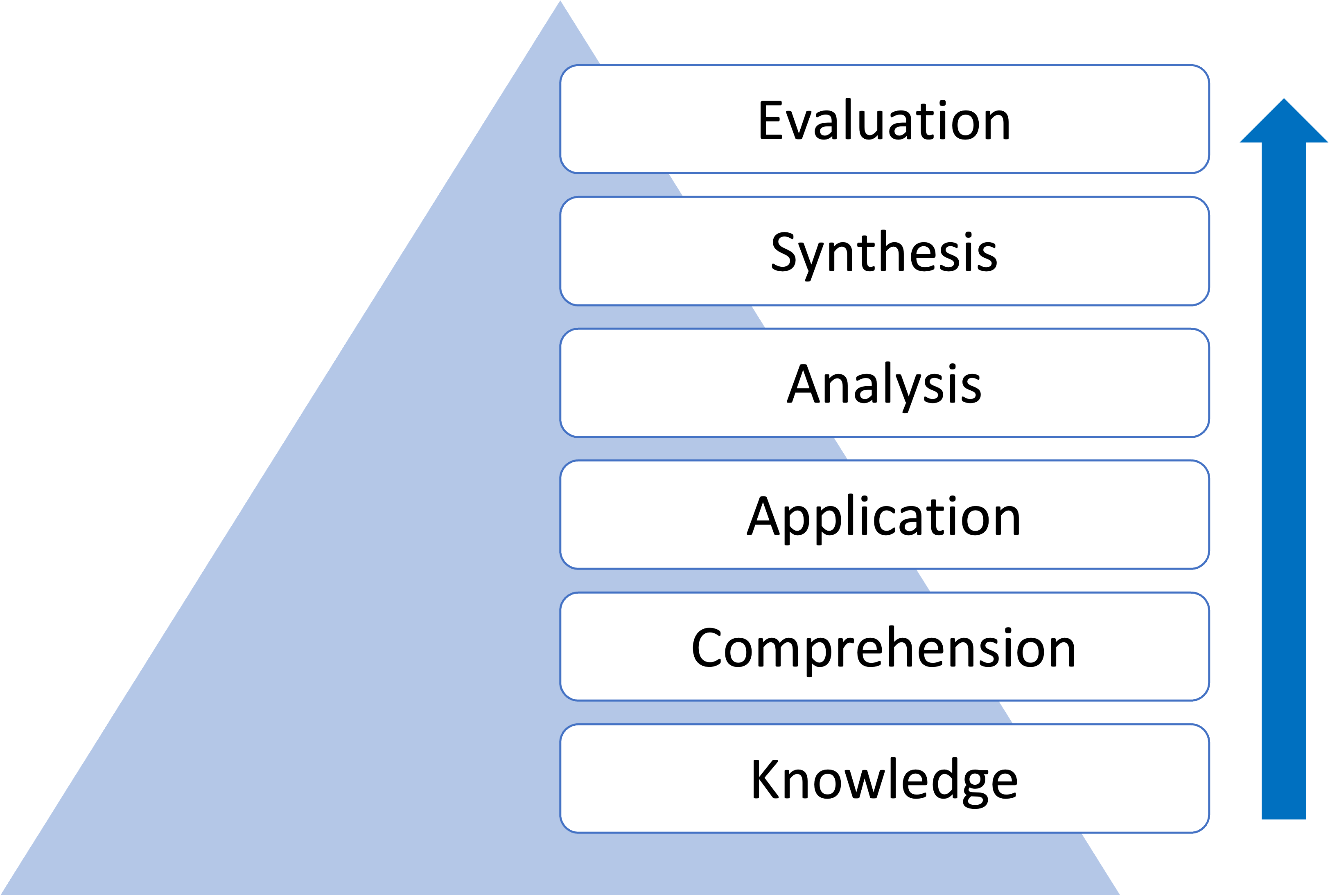} &
    \includegraphics[width=0.3\linewidth]{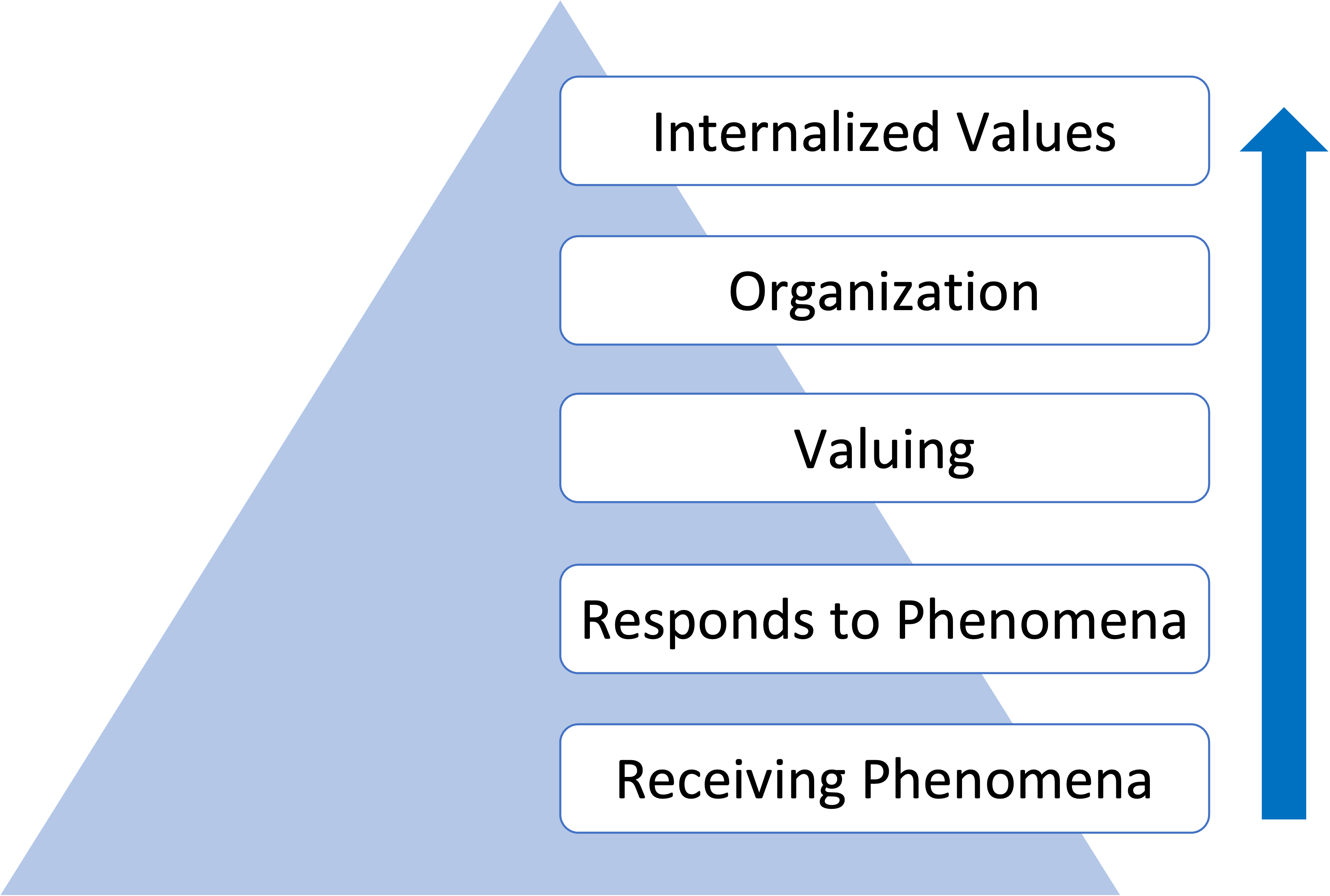} &
    \includegraphics[width=0.3\linewidth]{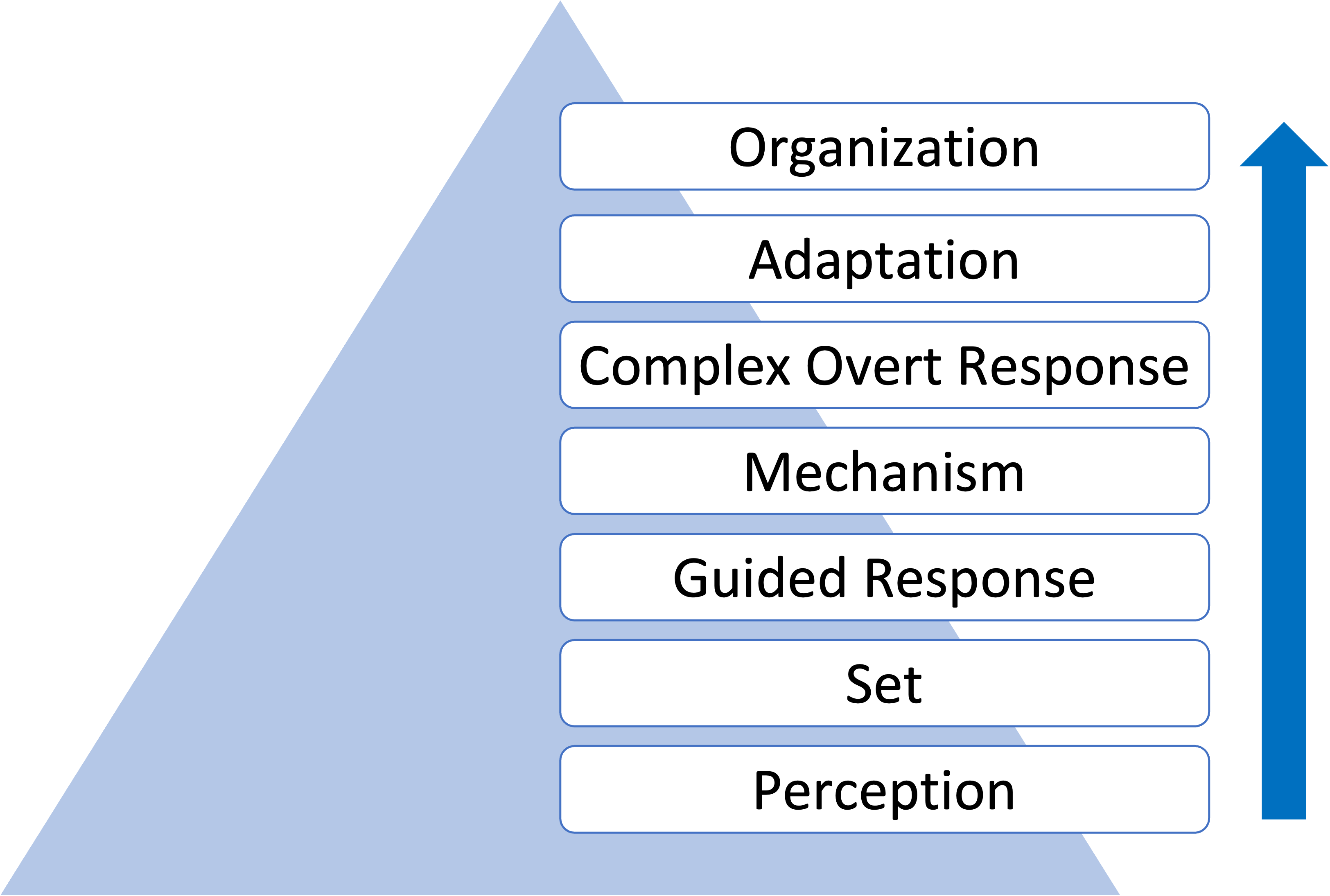}
    \end{tabular}
    \caption{Bloom's taxonomy domains: The cognitive domain (Knowledge-based), The Affective domain (Emotion-based), and The psychomotor domain (Action-based).}
    \label{bloom}
\end{table}

\begin{itemize}
    \item \textbf{Cognitive}: Related to mental skills and knowledge. This domain is consists of six major categories from basic to the most complicated mental skills including: Knowledge, Comprehension, Application, Analysis, Synpaper, and Evaluation.

    \item \textbf{Affective}: Related to behaviors or feeling developments. This domain is consists of five major categories from basic to the most complicated behaviors including: Receiving, Responding, Valuing, Organization, and Characterization.

    \item \textbf{Psychomotor}: Related to physical skills. This domain is consists of seven major categories from basic to the most complicated physical skills including: Perception, Set, Guided Response, Mechanism, Complex Overt Response, Adaptation, and Organization.

\end{itemize}

\begin{figure}[b]
	\begin{center}
	\includegraphics[width=0.8\linewidth]{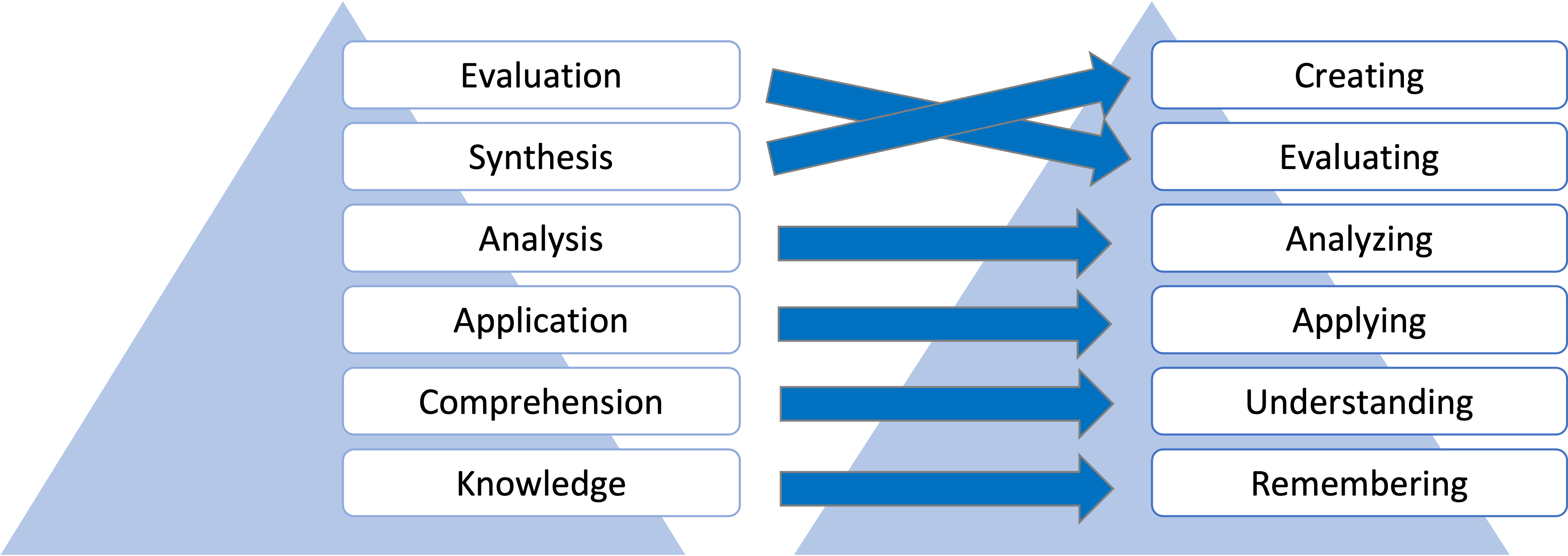}
	\end{center}
	\caption{A comparison between the cognitive domain of Bloom's taxonomy (left taxonomy) and Bloom's revised taxonomy (right taxonomy).}
	\label{bloom_compare}
\end{figure}

\textbf{Bloom's Revised Taxonomy.}
The Bloom's revised taxonomy is developed by Anderson and Krathwohl in mid-90s to promote the cognitive domain of the Bloom's taxonomy. They changed the names of the categories and rearranged their orders.
Figure~\ref{bloom_compare}
compares the cognitive domain in both taxonomies.

\subsubsection{A Taxonomy for Student Assessment}

\begin{figure}[t]
	\begin{center}
	\includegraphics[width=1\linewidth]{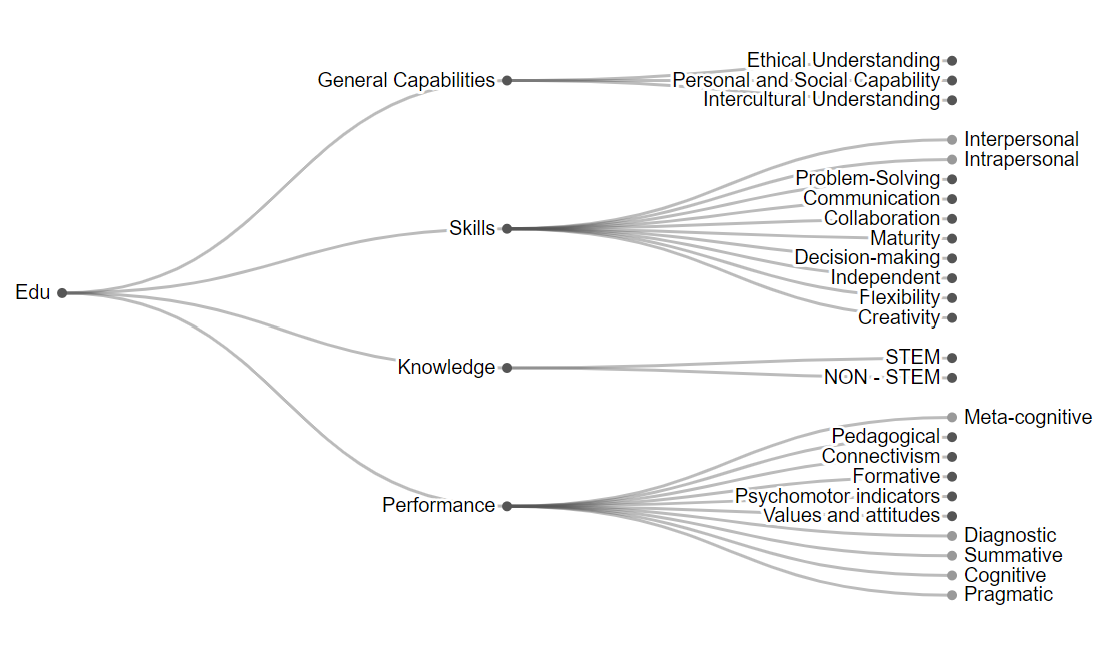}
	\end{center}
	\caption{Main concepts and sub-concepts of the education knowledge taxonomy for student assessment~\cite{schleicher2000measuring,loukusa2018assessing,wang2015assessment,schleicher1999measuring,wei2020assessment,balla1994assessment,jarchow2018measuring,auscurriculum}.}
	
	\label{assess_taxonomy}
\end{figure}

Student assessment is the process of assessing students' skills, abilities, and accomplishments. It is a constant, everyday part of the teaching and learning process in each and every classroom that is essential for teachers to evaluate their teaching techniques and students to understand the course material. Four main concepts that have been investigated in literature for students assessments are knowledge~\cite{schleicher2000measuring,loukusa2018assessing}, skills~\cite{wang2015assessment,schleicher1999measuring}, performance~\cite{wei2020assessment,balla1994assessment,jarchow2018measuring}, and general capabilities~\cite{auscurriculum}. The concepts and sub-concepts are all depicted in Figure~\ref{assess_taxonomy}.

\begin{itemize}
    \item \textbf{Knowledge}: Knowledge assessments are defined as measuring a student's academic strengths and weaknesses for understanding of academic subjects. Knowledge assessment is a kind of quantitative assessment, which is at the heart of education assessment. Teachers and parents use test scores to measure academic knowledge of students. The academic knowledge can be categorized to two main part of STEM and Non-STEM assessment~\cite{schleicher2000measuring,loukusa2018assessing}.
    \item \textbf{Skills}: Skill assessments measure the ability of students for doing an activity or task. Measuring students' skills can be helpful for understanding and increasing their styles of thinking and intellectual capability. Skills assessment will also evaluate the expertise of students in doing different tasks and how much they practiced them. Student skills can be categorized to interpersonal, intrapersonal, problem-solving, communication, collaboration, maturity, decision making, independent, flexibility, and creativity skills~\cite{wang2015assessment,schleicher1999measuring}.
    \item \textbf{Performance}: Performance assessment identifies the overall process of executing a task for students. It is a sequence of responses aimed at improving the environment in specified way. The sequence of responses consists of the behavior of a student, what a student says and dose and creates. Performance assessment defined as a response demand that focuses the students' achievement. Therefore, performance describes how a student behave or act by assessing pedagogical, connectivism, formative, psychomotor indicators, values and attitudes, diagnostic, summative, cognitive, meta-cognitive, and
    pragmatic behaviors~\cite{wei2020assessment,balla1994assessment}.
    \item \textbf{General Capabilities}: Capability in the Australian Curriculum encompasses knowledge, skills, behaviours, and attitudes. When studying in the classroom, it is crucial that students have the capacity to utilise their knowledge and talents in other, more difficult and ever-changing contexts. It is consists of ethical understanding, personal and social capability, and intercultural understanding~\cite{auscurriculum}.
\end{itemize}

\subsubsection{Creativity in Education}

Although creativity has always played a big role in human history, the present era due to the rapid change and its link with the concepts of innovation has become doubly important. So that creativity and invention have become the key measures of a country's success. With the increasing advancement of knowledge and technology and the widespread flow of information, societies today need training for creative thinking skills with which they can move forward with the developments. Hence, the goal should be to foster creativity in people from a young age through education.

Despite the recent research in creativity, it has been challenging for academics and researchers to consent on a definition to define it clearly. Generally, most of the definitions shares the two common attributes such as "novelty" and "effectiveness" (or usefulness). It is stated that an idea or product is creative if it is both novel (original) and useful for others. In other words, it should add something new and useful to the world that didn't exist before~\cite{amabile2011componential}. Third attribute is also considered in different researches such as "task appropriateness"~\cite{sternberg1999creativity}, "style of the product"~\cite{besemer1998creative}, and "wholeness"~\cite{henriksen2015novel} which means the aesthetic dimensions within specific context.

Hence, due to the most recent and clear definition of creativity, it is consist of three attributes including "novelty", "effectiveness", and "whole" or for short NEW. The NEW framework is then introduced to offer a systematic way to evaluate creative products in education. In fact, there are numerous instruments to evaluate creativity in education in literature such as divergent thinking tests, self-reported creative activities, judgments of products, and rating by others,
etc.~\cite{henriksen2015novel,allahbakhsh2013collusion,allahbakhsh2012reputation}.

The Torrance Tests of Creativity (TTCT) is one of the oldest and still popular method to evaluate creativity as a divergent thinking test method. The test contains two modes, figural and verbal, to be answered in a limited amount of time (30-40 min) then the scores from each subtests are combined into an overall score that covers four dimensions: Fluency, Elaboration, Originality, and Flexibility. Fluency refers to the amount of produced ideas; Originality refers to the amount of original and unique ideas; Elaboration is the capacity to elaborate on an idea and add details to it; and Flexibility is the capacity to generate a wide range of ideas.

Another recent research on creativity measurement is published by Baer J. et. al~\cite{baer2014gold} based on the rating method. The research is focused on the Consensual Assessment Technique which Teresa Amabile proposed it originally, and she and other scholars improved it afterwards. The writers believe that this method is a well-validated tool that has been called the gold standard of creativity assessment. It may be used to assess the originality of students' research ideas or scientific hypotheses, as well as their creative works and musical compositions, and their poems, essays, and stories. First, individuals will be asked to make anything (e.g., a short essay, poem, or experimental design) utilizing some basic instructions and materials. Then a group of domain experts will assess the creativity of the items they have created.

\textbf{The Components of Creativity.}
Based on Amabile's works on creativity, there are three components of creativity that determine a person's creativity including creative thinking skills, domain-relevant skills, and motivation ~\cite{amabile2011componential, amabile1998kill,amabile2012perspectives,hennessey1998reality, amabile2018creativity}.

The first component, creative thinking skill, depends on a person's cognitive style as well as personal characteristics. In general, people are more likely to be creative if they are identified by following cognitive styles: problem finding, idea generation, understanding complexity, using wide categories, remembering accurately, divergent thinking, suspend judgement, information organization, communication, spatial ability, and generate alternatives. This component also depends on people's personality like: self-discipline, tolerance for ambiguity, ability to delay gratification, high degree of autonomy, nonconformity, tolerance for risk taking, high level of self-initiated, tendency to overcome obstacles, curiosity, reflection, action, and social independence~\cite{amabile2011componential, james2001personality, guilford1973characteristics, barron1981creativity, sternberg1995defying,yakhchi2020enabling}.

The second component, domain-relevant skills, depends on people's education level, talent, and experience in a given domain. Expertise in a variety of interconnected disciplines might be helpful to become more creative. The third component, motivation, is called as the most important element of creativity since it determines what people will do with their expertise and creative thinking skills. There are two kinds of motivation, intrinsic and extrinsic. Intrinsic motivation like participating in an activity due to own interest, involvement, or personal challenge. And extrinsic motivation is related to environmental motivations like evaluative pressures, contracted for reward, surveillance, competition, and restricted choice.


\subsection{Educational Data Modeling }
\subsubsection{Data Modeling}

Data modeling is the process of building a data model to be used in an information system. An information system consists of a database which contains stored data and programs for collecting, manipulating, and recovering the data. Similarly, a data model describes the type of data to be stored and organized in the database in different formats~\cite{simsion2004data}.

There are many reasons to devote more time and resources to data modeling such as leverage, conciseness, and data quality. They all make a compelling case for considering the data model to be the most essential part of an information system~\cite{kruger2010data}.

\begin{itemize}
    \item \textbf{Leverage}: Programming may be made easier and less expensive with the use of a well-designed data model. Even a minor adjustment to the model can save a lot of money in terms of overall programming costs.
    \item \textbf{Conciseness}: The time it takes to evaluate a data model is far shorter than the time it takes to read hundreds of pages of functional specifications.
    \item \textbf{Data quality}: A database's data is generally a significant corporate asset that has been accumulated over time. Incorrect data lowers the asset's value and might be costly or difficult to fix.
\end{itemize}

\subsubsection{Data Modeling Terms}

Three basic terms and definitions for data modeling are an entity, attribute, and relationship~\cite{silberschatz1997database}.

\begin{itemize}
    \item \textbf{Entity}: An entity refers to a real-world object such as individuals, products, or organizations.
    \item \textbf{Attribute}: An attribute is a property of an entity such as age, color, or address.
    \item \textbf{Relationship}: A relationship is a connection between two entities.
\end{itemize}

\subsubsection{Types of Data Models}

Designing databases and information systems like any other system design start with a high degree of abstraction and gradually gets more concrete and particular. Based on the amount of granularity that they provide, data models can be divided into three categories. The procedure will begin with a conceptual model, then proceed to a logical model, and finally to a physical model~\cite{ballard1998data}.

Conceptual data models or domain models provide a broad overview of the system's contents, organization, and business concepts and rules. It is usually designed by data modelers and business specialists to obtain the initial requirements of the project design.

Logical data models provide more information and details on the concepts and their relations in the area being studied. The model determines how the system should run regardless of the database. It is commonly created by data modelers using DBMS and platform specifications to provide a technical map of data rules and structures.

Physical data models define a format or schema in which data will be stored in a database. This model is the most detailed of all that offers a finalized product to be implemented. It is mostly designed by database designers using performance requirements, DBMS and platform specifications, and the logical data model.

\subsubsection{Data Modeling Methods}

There are several data modeling methods to organize the data which are suitable for a specific data structure. The most well-known data models are~\cite{modell1992data, zhao1988object, mohamed2014relational}:

\begin{itemize}
    \item \textbf{Hierarchical Data Model}: This approach is well suited to situations when the information collection is based on an actual hierarchy. The model is a tree shape or parent-child hierarchical structure. The schema contains a parent root record and several child. records or subdirectory branches. The root record links to the child records and each child record can be linked to several additional subdirectories.
    The hierarchical model have been used widely in education, e.g., to measure educational service quality~\cite{teeroovengadum2016measuring}, and to evaluate extrinsic and intrinsic motivation in students~\cite{vallerand2002intrinsic}.
    \item \textbf{Network Data Model}: This approach enhances the hierarchical data model through enabling the existence of numerous parent records which means allowing each child record to be linked to several parent records. In education, the network model have been used to, e.g., emphasize on the importance of education in environmental protection~\cite{yu2020construction}, develop a learning network model for higher education consortia formation and management~\cite{sagan1969network}.~\cite{sweet2013hierarchical} also presented a new framework called Hierarchical Network Models (HNM) for educational research and developed single-network statistical network models to multiple networks.
    \item \textbf{Object-oriented Data Model(OODM)}: This approach is based on object-oriented programming and assists designers to represent the information with an expressive tool and facilitating the access of users to the information. It is typically employed in applications that seek for high performance, computations, and quick outcomes.
    \item \textbf{Relational Data Model}: A relational data model consists of a set of tables, recognized as relations, are consists of rows and columns. Every row, or tuple, represents a single instance of the entity. Every column corresponds to an entity's attribute. A domain is the name given to all of the attributes of a relation. There are also different types of connections between the tables such as one to one, one to many, and many to many that the model takes into consideration. The Structured Query Language (SQL) as a standard language is commonly used to create a relational database to access, use, and modify databases.
    This method is the most used data model in education, e.g.,~\cite{bogdanovic2008development} has introduced a tool that simplified and partially automated the process of designing relational educational data for students, and~\cite{winer1990use} examined employing relational model as a data analysis and  management tool to study educational environments.
    \item \textbf{NoSQL Data Model}: Other non-relational or non-SQL models have been developed such as document model, multivalue model, and graph data model~\cite{beheshti2012framework}. These three are prominent examples of NoSQL or not only SQL data model.

    \begin{itemize}
        \item The Document Model stores and manages semi-structured data or documents instead of atomic data~\cite{mason2015nosql}. For instance, the background educational data gathered from students' activities using a document model proved to be helpful to create adaptable educational documents~\cite{franze1999document}.
        \item The Multivalue Model allows the attributes to take a list of data instead of a single point which makes it different from the relational data model. In other words, in a multivalue model an attribute (cell) may contain an array of values. In education, this model proved to be usefull to make the process of data analysis faster by using multidimensional arrays of student values~\cite{martin2007examining}.
        \item The Graph Data Model allows any node connection with different structures coming from various sources of information~\cite{frisendal2016graph}. In the next subsection, we will explain the terms and concepts related to this model .
    \end{itemize}
\end{itemize}

\subsubsection{Graph Data Modeling}
\label{knowledge_graph_RW}
\textbf{Graph data.}
Graphs have recently become a popular data structure for data analysis because of their capability to depict any type of real-life problem.

As you can see in Figure~\ref{graph}.a the structure is composed of the following: node or vertice, edge or link, and property. Node as the basic component of a graph structure represents entity. The connection between nodes is called as an edge which represents relation. And property as attribute provides more information about the nodes and edges (Figure~\ref{graph}.b)~\cite{bondy2008graduate, patel2017graph}.

\textbf{Knowledge Graph.}
Graph-based visualizations are able to address a wide range of topics from several fields, including natural language processing (NLP), knowledge discovery, large network systems, social networks, etc~\cite{velampalli2017graph}. Knowledge Graph (KG) is distinguished from other knowledge-based information systems by their unique integration of knowledge visualization system, algorithmic searches, and information management procedures~\cite{gomez2017enterprise}.

Although there are several different definition of KG in the literature, an integrated formulated definition is provided by~\cite{rizun2019knowledge} as follows:

\begin{figure}[t]
	\begin{center}
	\includegraphics[width=0.9\linewidth]{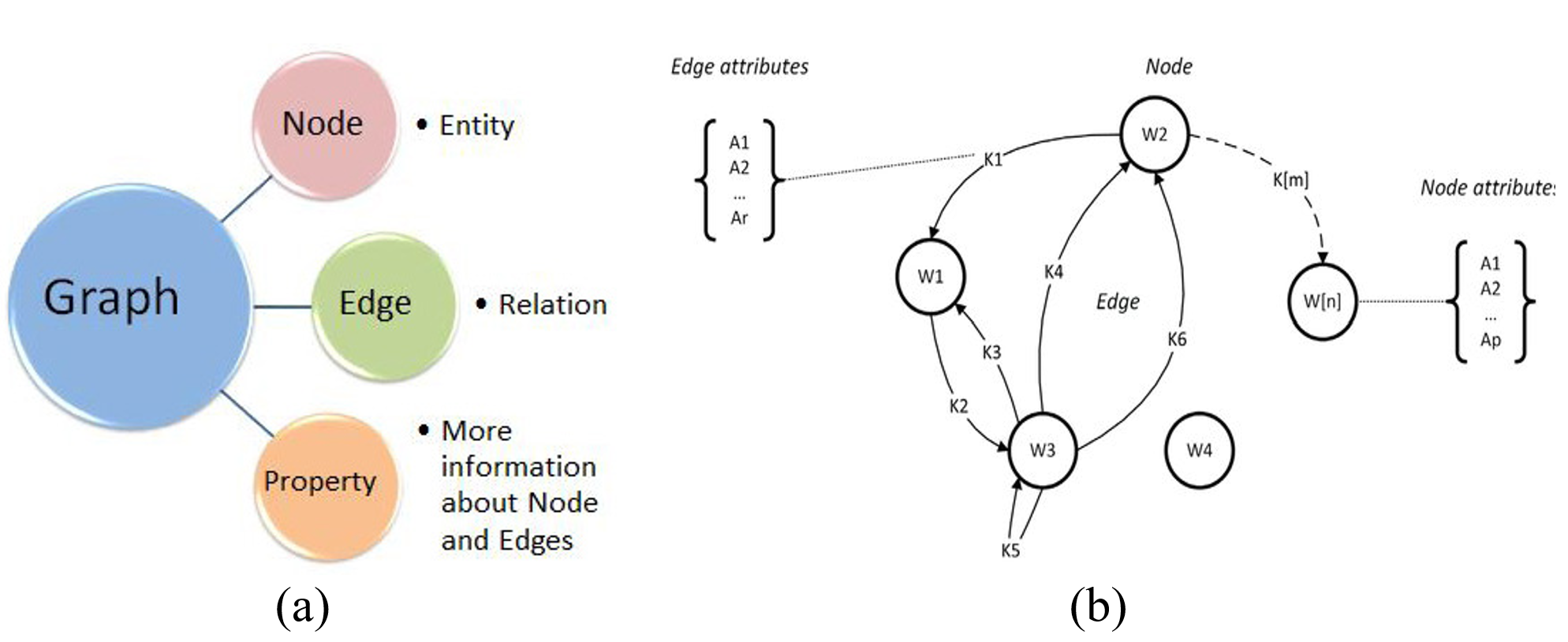}
	\end{center}
	\caption{(a) Graph structure. (b) Graph components ~\cite{patel2017graph, czerepicki2016application}.}
	\label{graph}
\end{figure}

\begin{displayquote}
``A Knowledge Graph is a knowledge base that (1) replicates the model of information flow in an organization, (2) stores complex structured and unstructured knowledge, (3) is presented in the form of entities and relations between them, (4) covers a multitude of topical domains, (5) acquires and integrates knowledge, and (6) enables interrelation of arbitrary entities.''
\end{displayquote}

\begin{figure}[t]
	\begin{center}
	\includegraphics[width=0.5\linewidth]{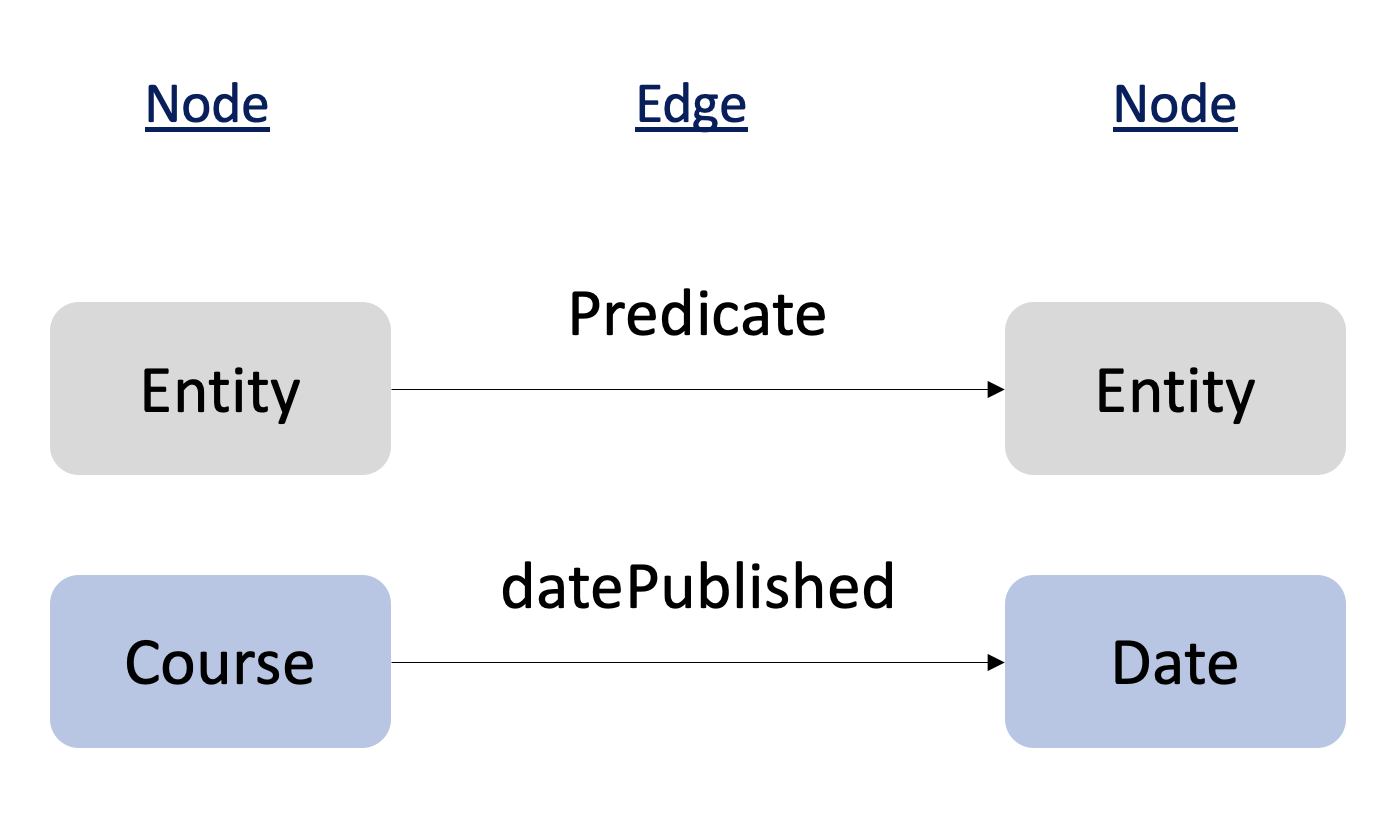}
	\end{center}
	\caption{The structure of a Triple and an example of it.}
	\label{triple}
\end{figure}

KG model is a database that stores information according to a Knowledge Schema (KS). KS supplies the KG's basic structure and build its metalayer, as well as defining containers as classes for entities of comparable types. Posting and interconnecting data on the Web utilizing the Resource Description Framework (RDF) is one of the approaches of keeping information in a Schema that has lately gained popularity. According to the RDF standard developed by the Word Wide Web Consortuim (W3C) in 2019 knowledge can be depicted using Triples. The Triples include two entities as head and tail and a predicate which connects them~\cite{wu2017knowledge}. Figure~\ref{triple} depict a Triple with two entities as "Course" and "Date" and a predicate as "datePublished".

KG has been investigated in depth and used in a variety of fields of science with controversial ideas. Based on various research questions stated in different resources~\cite{socher2013reasoning,wu2017knowledge,tian2016learning}, it can be concluded that Knowledge Graph Embedding is the key concept of KG.  It is the procedure of converting graph components (e.g., entities and relations) into higher-dimensional vectors in order to make data operation easier.

In this regard, different embedding methods have been introduced, e.g., the RESCAL paradigm which uses multirelational data to accomplish collective learning~\cite{nickel2011three}, the Skip-gram model as an unsupervised model which predicts a contextual word using a target word making the process of text data easier~\cite{mikolov2013efficient}, the TRESCAL model as an improvement to  RESCAL model performs relation extraction~\cite{chang2014typed,beheshti2017systematic}, and TransE which transforms entities and relations of a graph to vectors in the same semantic dimension space~\cite{bordes2013translating}.

\subsection{Open Learner Models}

A learner model represents the knowledge, misunderstandings, and challenges of learners~\cite{bull2004open}. Learner models are often concealed from the learner they are representing, as they are created from the data acquired from advanced educational environments to adjust to a user's learning requirements. Currently, systems are more inclined to interact with their learners and allow them to see and assess their model using an Open Learner Model (OLM). OLMs propose a new paradigm for revealing the model to students, making it more helpful to them. Such learner models encourage metacognitive skills such as self-awareness and self-regulation by allowing students to analyze their reflections.

According to Zimmerman~\cite{zimmerman2002becoming}, the ability of students to recognize and manage their educational environment, including internal and external dynamics, is referred to as self-regulated learning. A self-regulated learner is also defined as the way they are actively engaged in their education in a cognitive, metacognitive, and motivational manner. Furthermore, research reveals that technological approaches which augment the thinking skills and self-regulation of students will improve their performance and learning outcomes in all areas~\cite{hooshyar2019potential,beheshti2020towards}.

Regarding this, OLM can be a tool to support self-regulated learning as it helps the learners to view their information. Thus, learners will be in a position to observe, evaluate, and manage their learning which will enhance their educational outcomes by changing their learning strategy as necessary~\cite{bull2010open}. In other words, the model is beneficial for both teacher and learner as it provides personalized learning based on the learner's progress and behavior.

Other than supporting metacognitive skills, many other goals are identified by (Bull, 2007) for developing OLMs and then designed a framework for OLM refferred to as SMILI:) to answer three questions including, "what is available?", "How is the model presented", and "Who controls access?"~\cite{bull2007student}. The purposes of the model are defined to improve the accuracy of learner model, encourage learners to reflect, assisting learners by keeping track of their progress, promote competition and cooperation among learners, permit the learner to have access to their data, assist with navigation, assist with learner control, improve student confidence in the system by displaying learner model content, and assess by using the learner model.

\subsubsection{Representation of OLM}

There are two kind of representation approaches for visualizing models, simple and complex, which basically depends on the purpose of developing and the users of the OLM. Simple models that just represent the knowledge level of a learner in a variety of domains can only provide the learner with this basic information. They basically employ a number of graphical forms to convey a series of concepts such as using graphs, charts, boxes, tables, or skill meters to depict the knowledge and progress level of a learner. 

\begin{figure}[t]
	\begin{center}
	\includegraphics[width=0.7\linewidth]{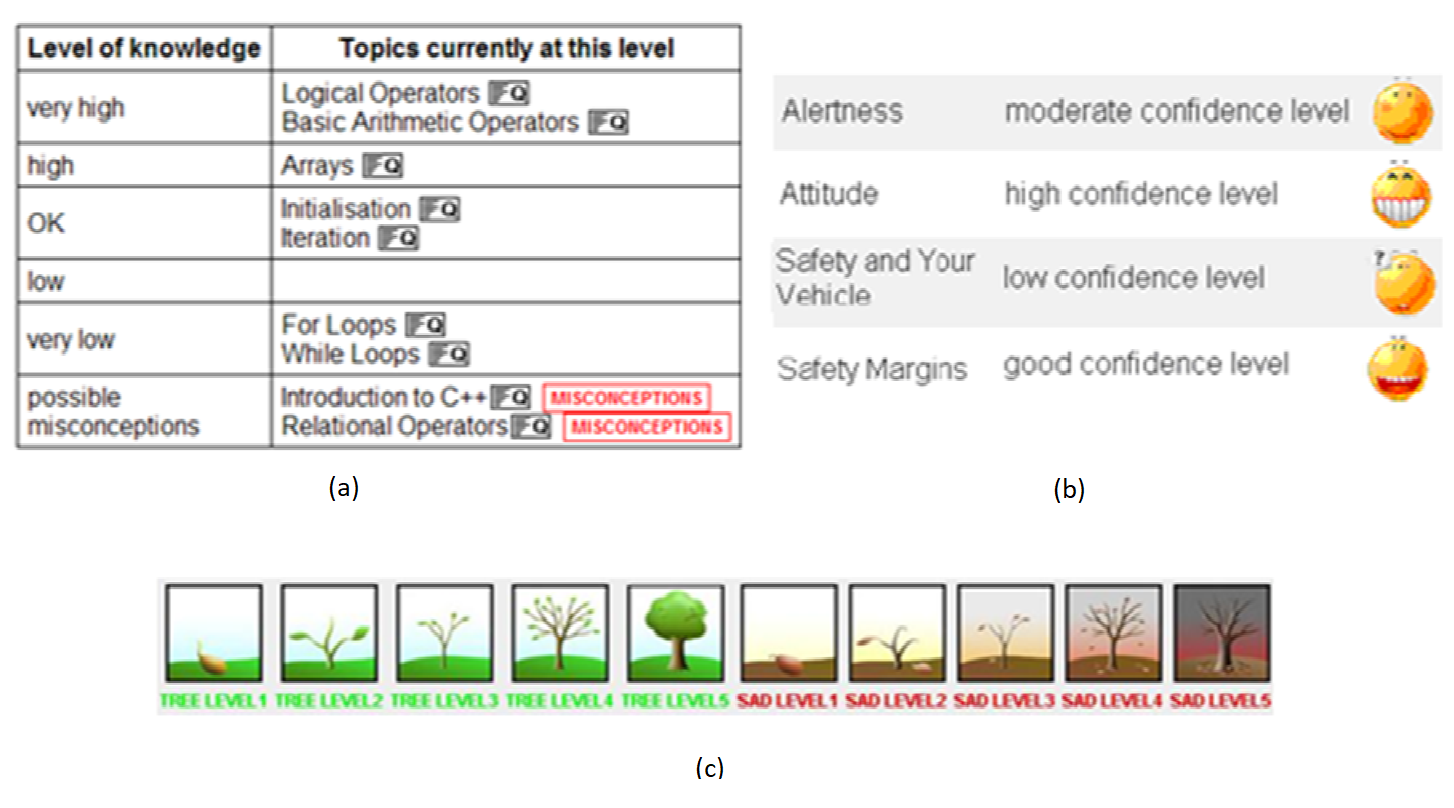}
	\end{center}
	\caption{Simple OLM format examples using: (a) ranked list, (b) smiley faces, (c) growth of trees~\cite{ahmad2014users}.}
	\label{simple_OLM}
\end{figure}

Simple graphs may represent learners in a simple numerical model or use more complex formats such as a combination of skill meters and a structural view. There are also other popular formats to display the same skill meters like using pictures of tree growth~\cite{lee2008open}, the liquid level in a container~\cite{papanikolaou2003personalizing}, smiley faces~\cite{kerly2008children}, and arrows in a target~\cite{brusilovsky2005user}. Figure~\ref{simple_OLM} depicts some of these skill meters which are more appropriate to be used for children that may encourage them to expand their knowledge.

Complex or structured models provide learners with more comprehensive information and more complicated knowledge structured views. But it does not imply that the learner has access to all the information available in the model, accessing to the information will be defined for each user and learner by an administrator~\cite{hooshyar2019potential}. Like simple models, they also employ a variety of visualization techniques such as tree maps, three dimensional structure, hierarchical tree structures, textual description, and concept maps. While the most common visualization method in simple view is skill meter,  the most common structure in complex view is concept map. 

\subsubsection{Types of OLM}

Providing learners some control over their education can motivate them to become more self-reliant and self-regulated. OLMs offer various learner control such as Editable, Co-operative, Inspectable, and Negotiated types \cite {hamzah2018learner}.

\textbf{Editable} OLM allows learners to modify the data that the system provides. This interaction is allowed due to reasons such as expanding knowledge, and recalling information or resources. Hence, learners are fully responsible for their learner models since they have the ability to modify the information as their knowledge evolves. For example, this method is used by the Flexi-OLM system~\cite{mabbott2006student}, the C-POLMILE system~\cite{bull2003learner}, and Scrutable Adaptive Hypertext system~\cite{czarkowski2005web}.

\textbf{Co-operative} OLM provides some means of integrating the views of the tutor and the learner to construct a more comprehensive model rather than completely relying on the system to create the model. Since both sources of input should be combined in a process, building this model is more complicated. Mr. Collins ~\cite{bull1995did} is an example of a cooperative OLM under the control of both tutors and learners. In this model, the learner can challenge the system's assessment and if the tutor disagrees with the claim, the system would challenge the learner with a question concerning the disputed topic.

\textbf{Inspectable} OLM offers read-only data to the learners and is solely under control of the system. The model is completely reliant upon system deduction which is derived from engagements between the system and learner. The learner may view the model but not alter its content unless they use the standard methods like asking the learner to answer more questions~\cite{bull2010open}.

\textbf{Negotiated} OLM is the product of a process of negotiation between learners and the system to come up with a model that both the learner and the system agree on. The procedure often include seeking and offering information, validating, challenging, arguing, and accepting. Mr. Collins~\cite{bull1995did} is one of the first negotiated OLMs which is based on involvements among learner and the system model. In this model, the contrasts in beliefs are clearly reflected. In comparison, STyLE-OLM~\cite{dimitrova2003style} only represent one belief which both the system and learner consent. In the end, any disputes that emerge will be addressed through debate or eliminated, and the agreement achieved will be included to the model. Many methods have been used to perform the negotiations including \textit{dialogue game} (used in STyLE-OLM)~\cite{dimitrova2003style}, \textit{menu selection} (used in Mr. Collins)~\cite{bull1995did}, and \textit{chatbots} (used in CALMsystem)~\cite{kerly2007calmsystem}.

\subsubsection{OLM Supporting Creativity in Learners}

There are also a number of research specifically addressing creativity and its related skills in OLMs~\cite{clayphan2013open, clayphan2016wild, papanikolaou2014constructing,bull2008metacognition}. For example, Clayphan et. all~\cite{clayphan2013open} created the first OLM that supports group and individual brainstorming reflections and claim that the model is effective and can enhance creativity in learners. They have designed a model asking learners three main questions, perform several brainstorming sessions, and analyze the answers to build the learner model. The questions are as follows: "How much did I contribute?", "At what time was the group or an individual stuck?", and "Where did group members seem to spark off each other?".

\begin{figure}[t]
	\begin{center}
	\includegraphics[width=0.7\linewidth]{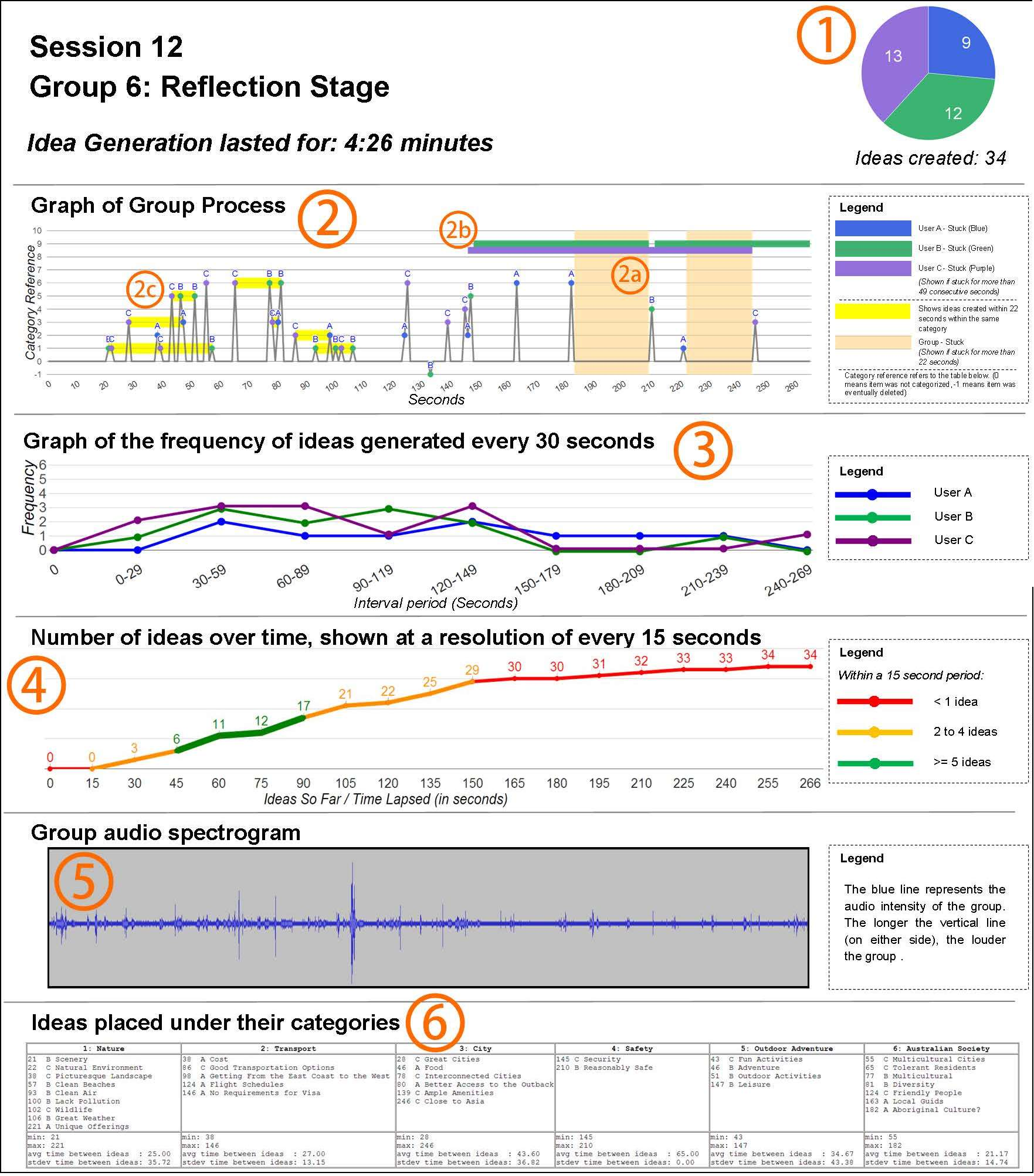}
	\end{center}
	\caption{An OLM visualization supporting brainstorming reflections~\cite{clayphan2013open}.}
	\label{creativity_OLM}
\end{figure}

Figure~\ref{creativity_OLM} depicts one of the sessions of the learner mode that assists learners to answer the mentioned questions. The design has 6 parts that are all marked in the figure:
(1) Each person's total amount of ideas.
(2) Graph of group process. In this part, the horizontal and vertical axes show time and category reference, respectively. Each dot is an idea and their color indicate the learner which is defines in the legend (e.g., here blue color indicate user A). The Orange rectangles indicate the amount of time groups was stuck for more than 22 seconds (2a). The Bars show the amount of time each user (A, B, or C) was stuck for more that 49 continuous seconds (2b). And, the yellow bars show ideas created within 22 seconds in the same categories (2c).
(3) Graph of the frequency of ideas generated every 30 seconds by each person.
(4) Number of ideas over time, shown at a resolution of every 15 seconds, cumulatively. (5) Group audio spectrogram which is useful when group is stuck and need discussions. (6) The list of all the ideas placed under their categories based on user and time of creation.

To evaluate the learner model they have invited 15 participates to perform a thinking aloud study and asked them to answer a sequence of questions in three anonymized brainstorming sessions. Results show that the model is successful to help participants evaluate their own performance by answering a series of questions using interactive tabletops. This gives them the opportunity to find out more about their contribution level, when they got stuck, and when they lit up each other with ideas in a group.

\subsection{Educational Data Mining and Learning Analytics}
\subsubsection{Data Mining}

The fast advancement of technology has resulted in a massive influx of data from a variety of sources. The rise in the volume of data, as well as the development of new technologies, has created a serious concern. Data gathering has also risen significantly, and traditional software's ability to acquire, handle and analyze such massive amounts of data has become obsolete. Massive amounts of data are now available all over the world, requiring analysis to extract potentially valuable and previously uncovered information which the traditional data analysis techniques are unable to deliver~\cite{wu2013data}.

To produce the desired demands, Data Mining (DM) originated as a response to the need to create new data analysis methodologies. DM has become a popular study topic as a result of the growing demand from social media and the practical side of things. This new branch of computer science examines historical data to extract previously undiscovered and beneficial knowledge for decision-making. It detects patterns that aren't visible to the naked eye. If there is a big volume of data accessible for analysis, the quality of the derived knowledge will be higher. Artificial intelligence, data visualization, database technology and machine learning are all components of DM as a multidisciplinary field of science ~\cite{salma2021DM,amouzgar2018isheets}.

DM, the practice of extracting patterns from data, is a process that should be automatic or semi-automatic and the discovered patterns should indicative in a sense that it provides some advantage~\cite{witten2005practical}. It isn't restricted to a single form of data and it can be implemented on unstructured data (e.g., documents), semi-structured (e.g., XML-based), and structured (e.g., tables in Relational Databases) datasets~\cite{han2011DM}.

\textbf{DM Techniques.}
DM learning techniques are classified as supervised, unsupervised, and semi-supervised learning. In supervised learning, it is required all data to be labeled and during the training phase, the system is given a pair of inputs (X) and outputs (Y) to predict a specific quantity. Differently, no labeled data is required in unsupervised learning, and the system is merely provided input (X). The main goal in unsupervised learning is to understand the data and look for structure or unusual patterns in data analysis. Semi-supervised learning integrates techniques from both supervised and unsupervised learning since some of the data is labeled while the remaining is not~\cite{liao2012data}.

Descriptive and predictive models are two different types of DM models. Descriptive models detect similar items or common groups in existing data, with the aim of identifying reasons for success or failure while predictive models categorize events in the future in a deeper way and try to satisfy the unknown results in advance~\cite{pujari2001data}.

A descriptive model presents the main features of the dataset. It simply provides a summary of the data points that allow users to look at specific elements of the dataset~\cite{ghodratnama2021intelligent}. A descriptive model is often discovered by unsupervised data mining to identify patterns in the datasets, but it is up to the data miner to analyze the discovered patterns~\cite{olson2019descriptive}.  Examples of the descriptive model are Clustering (grouping similar records)~\cite{rajagopal2011customer}, Association Rule (Identifying relationships between records)~\cite{ solanki2015survey}, Sequence Discovery (discovers statistically relevant patterns in sequential data)~\cite{lobb2016novel}, and Summarization (finding a compact description of a dataset)~\cite{ertoz2003detection}.

Predicting the value of a new unknown variable, the target variable, is the goal of a predictive model. The data mining task is termed Classification if the goal value is one of a specified number of defined class labels~\cite{phyu2009survey}. If the target variable is continuous, the task is Regression~\cite{gupta2015regression}. Prediction~\cite{weber2009data}, and Time Series Analysis~\cite{esling2012time} are two other prominent techniques that are using in DM. The input variables for each form of prediction might be categorical or continuous; based on the type of input data utilized, different prediction approaches are more likely to be efficient~\cite{baker2010data}.
Top three data mining techniques are explained below:

\begin{itemize}
    \item \textbf{Classification} as a supervised learning technique classifies data types or concepts to predefined classes. In data mining, classification is a well-known important study field, and enough research has been done in recent years to enhance key problem strategies. In classification, the core concepts is to make predictions depending on the user's interests and desires. Data can be classified by various criteria such as the type of mined data (e.g., text, spatial, multimedia), the data model (e.g., relational database), type of discovered knowledge (e.g., discrimination, characterization), or the data analysis approach (e.g., Neural Networks, statistics, machine learning)~\cite{phyu2009survey,kausar20213d,jalayer2022ham,niu2020deep}.

    \item \textbf{Clustering} as an un-supervised learning technique clusters data based on shared characteristics. In other words, those objects that are similar to each other are grouped in one category and the rest are grouped in another category. Clustering methods include Hierarchical, Partitional, and Densitybased clustering~\cite{Berkhin2006}. Hierarchical clustering creates a hierarchy of nested clusters by repeatedly executing some criteria. Partitional clustering classifies a given dataset into multiple groups using some criteria. And, density-based clustering groups the given data based on the density of points in data space. It works based on the notion that an adjacent region of high point density defines a cluster~\cite{Sander2010}. 

    \item \textbf{Regression} is a supervised learning technique that has two types: Linear and multiple regression models.
    \begin{enumerate}
        \item Linear Regression: The most common application of this type is to employ a linear equation to represent the relationship between two variables. It may also be utilized to determine the mathematical connection between the variables. It is the most basic type of regression and the model's formula is as follows:
        \[ A + Xd = Y \]
        \item Multiple Regression: This model as one of the most often used models in data mining for making predictions is most commonly used to describe the connection between several independent or predictive factors. Generally, the model predicts an outcome using two or more independent factors. The following is the formula for the multiple regression model:
        \[ Y = a_0 + a_1x^1 + a_2x^2 + a_3x^3 + ... + a_kx^k + e\]
        In data mining, classification and regression are two of the most common prediction techniques. Since they are so similar, it might be challenging for a user to know whether to use one over the other. 
    \end{enumerate}

\end{itemize}

\subsubsection{Educational Data Mining}

Recently, Educational Data Mining (EDM) as an application of DM in education has emerged to improve teaching, learning, and research outcomes. At first, it was introduced by a few workshops and seminars in Artificial Intelligence in Education (AIED) conferences and now it has progressed to the point where it now has its community and societies. 

On the subject, there is a substantial and varied body of literature. For example, Bienkowski et al. (2012)~\cite{bienkowski2012enhancing} offers a widely referenced study in which they introduce EDM, as well as its basis, difficulties, and implications. There are also several books and paper surveys that present applicability, approaches, and the state of the art in EDM~\cite{romero2006data, romero2010handbook, pena2014educational,romero2014survey,romero2020educational}.

\begin{figure}[t]
	\begin{center}
	\includegraphics[width=1\linewidth]{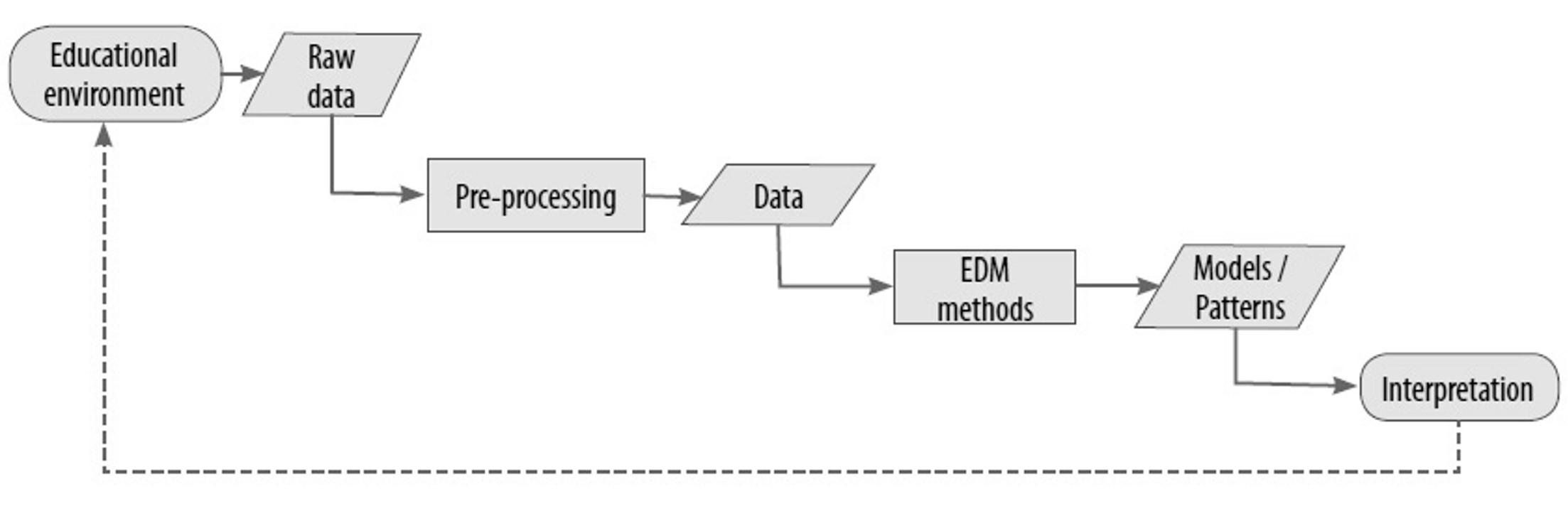}
	\end{center}
	\caption{Overview of applying EDM methods~\cite{linan2015educational}.}
	\label{edm_process}
\end{figure}

The application of EDM techniques includes a number of processes. As you can see in  Figure~\ref{edm_process} the depicted steps in this process are as follows~\cite{linan2015educational}: (1) Designing a plan and identifying the necessary data, (2) Extracting the data from available educational environments, (3) Curating the extracted data since it might originate from range of sources with different formats, (4) Applying EDM methods to create models/patterns, and (5) Interpreting the models/patterns. In case of inconclusive outcomes, after making changes to the teaching/learning method or the study design, the analysis will be repeated. This might happens because of not properly addressing the problem, inadequate or unsuitable input raw data, or choosing not powerful methods.

There are four areas of EDM applications that are increasingly popular within the field~\cite{baker2010datamining}:

\begin{enumerate}
    \item \textbf{Enhancing student models}: The educational data about students' charcterisitc or behavior, e.g. their cognitive, meta-cognition, motivation, and knowledge, will be used to personalized the learning process and modeling differences between students~\cite{wang2021assessment2vec}.
    \item \textbf{Domain knowledge structure models}: The EDM methods that are a combination of psychometric modeling framework and advanced space-searching algorithm are designed to discover accurate domain models directly from data.
    \item \textbf{Research on pedagogical supports}: Learning software provide various kind of pedagogical support for students. Identifying the most efficient educational support is another key application of EDM.
    \item \textbf{Research on learners and learning}: Research on questions in any of the above mentioned fields (e.g., student modeling, domain modeling, and pedagogical support) may have larger scientific implications for learning and specifically for the learners in the system.
\end{enumerate}

\subsubsection{Learning Analytics}

There is another related community called Learning Analytics (LA) which shares a common interest with EDM in the way educational data
has the potential to make a difference in terms of learning outcomes. It initially defined as the Learning analytics is the measurement, collection, analysis, and reporting of data about learners and their contexts, for the purposes of understanding and optimizing learning and the environments in which it occurs. in the first International Conference on LA, also held in Canada in 2011~\cite{siemens2013learning}.

There is a substantial and diverse collection of literature on LA. For, example, Larusson (2014)~\cite{larusson2014learning} as a popular book in LA contains the most up-to-date concepts, research, methods, techniques, and case studies. It helps the readers find out how to: (a) improve the performance of students and educators; (b) increase student comprehension of course content; (c) Identify and respond to the needs of students who need help; (d) enhance grading accuracy; (e) enable institutions to make better use of their resources; and (f) allow educators to evaluate and improve their own abilities.

The application process of LA is similar to EDM and the steps for testing a hypopaper linked to the learning/teaching process are the same as for EDM. It consists of data extraction from relevant educational enviroments, preprocessing extracted data, applying numerical/computational methods to help educators or psychologist to interpret the outcomes and make decisions.

\subsubsection{Common Methods in EDM and LA}

The communities of EDM and LA are both interested in exploring new methods to enhance the research on big educational data. Also, there are differences in terms of the research questions, methods and techniques, and focuses, the two areas have a lot in common both in goals and the methodologies and strategies employed throughout the research~\cite{baker2014educational}. They can be seen as an intersection of three areas: Education, Statistics, and Computer Science.

The EDM/LA process is a cycle of data mining and knowledge discovery which is involved with students, instructors or academic authorities, and educational environments to produce educational data and ultimately new knowledge. This new knowledge is gained by EDM and LA methods and will be used again by students and instructors in the process~\cite{romero2020educational}. The majority of methods pertinent to educational data are used both in EDM and LA. The most popular methods are (1) clustering, (2) prediction, (3) Social network analysis, (4) outlier detection, (5) casual mining, (6) process mining, (7) relationship mining, and (8) text mining. There are also prominent researches around discovery with models and distillation of data for human judgment~\cite{bakhshinategh2018educational, siemens2012learning,siemens2013learning,beheshti2022social}.

\begin{itemize}
    \item \textbf{Prediction} refers to estimating unknown variables based on history data of a mixture of additional variables. In education the variables might represent performance of students, their characteristics, or learning outcomes~\cite{bakhshinategh2018educational}. The majority of extant literature attempts to estimate academic achievements of students, although studies also explore students' characteristics such as social interaction~\cite{raju2015exploring}.
    \item \textbf{Clustering} refers to dividing data point into categories based on similar observations. It is different from classification since the class labels are not defined. Hence, it is sometimes called unsupervised classification. For instance, clustering method has been used to group schools based on their common features in education~\cite{beal2006classifying}.
    \item \textbf{Social network analysis or SNA} refers to studying the social relationship between individuals in a network. To Investigate online social interactions, data mining methods (e.g., classification, prediction) can be used on the network data. SNA can also be used on education data to mine students' group activities~\cite{ reyes2005mining}.
    \item \textbf{Outlier detection} refers to detecting data points that are considerably different from the majority of data. In education, outlier detection has been used to identify students who have learning difficulties or who have unusual learning patterns~\cite{ueno2003line}. It has also been used to detect irregular behaviors and interaction between educators and students~\cite{oskouei2013perceived}.
    \item \textbf{Causal mining} refers to seeking for causal relationship in data. In education, it has been used to identify the characteristics of students' behaviour that contribute to their learning outcomes, problems, grades, etc~\cite{ de2019causality}.
    Classification and Regression are two widely used prediction methods that have been used for estimating student's performance and characteristics~\cite{ahmad2015prediction,el2019multiple}.
    \item \textbf{Process mining} refers to obtaining information from the process history logs. It can used to trace the evolution of student behavior through using educational environments.~\cite{ juhavnak2019using} has used process mining method to evaluate the behavior patterns of students while taking quizzes.
    \item \textbf{Relationship mining} refers to discovering relationships between variables within a big database~\cite{rajabi2016interlinking}. It is involved with identifying the factors that are most closely linked to a certain variable of interest. Associate rule mining~\cite{tabebordbar2018adaptive,tabebordbar2020feature,tabebordbar2020adaptive}, frequent pattern mining, and sequential rule mining are three common types of relationship mining method.
    For instance, associate rule mining as the most common method is an if-then rule aiming to see if one action affects another by looking at the representation of actions in the dataset. As an example, eLORM is developed to mine the Learning Objective relationships centered on the student's use information~\cite{ouyang2007elorm}.
    \item \textbf{Text mining} refers to mining knowledge from text. In education, text mining has been used to analyze the contents of the students' discussion forums, chats, websites, emails, or any text document~\cite{reategui2011sobek}. This method has also been used for the purpose of educational document classification based on their similarities and also checking plagiarism~\cite{oberreuter2013text}.
    \item \textbf{Discovery with models} is the process of including a prior verified model of a phenomena into a new analysis. In discovery, human reasoning, instead of automated approaches, is used to build models based on prediction, clustering, or knowledge engineering. Then, the created model is integrated into larger models, such as relationship mining. For example,~\cite{ hershkovitz2013discovery} has used this method to investigate on the relationship between carelessness measures and students' intended goals.
    \item \textbf{Distillation of data for human judgment} refers to using distillation, interactive platforms, and visualization to display data in an understandable manner. In education, this method has been used for two main reasons: identification and/or classification. The goal of distillation for identification is providing data in a way that it can be identified using recognized patterns. Using data distillation to represent the states of students' knowledge as they learn new skills is an application~\cite{ corbett1994knowledge}. Distillation of data for classification can also be used as pre-processing step before creating a prediction model, e.g.~\cite{d2008more} has used this method for the purpose of knowledge tracing in students.

\end{itemize}

\subsubsection{Creativity Assessment Using EDM/LA}

Although much have been investigated regarding EDM and LA in education and student assessment, there is little study on creativity assessment using EDM or LA. However, there are some researches around cognitive processes such as creative thinking and problem-solving skills~\cite{deng2018innovative,lu2010using,singelmann2020innovators, yu2010rough,barukh2021cognitive}.
For example, Yu~\cite{ yu2010rough} performed rule extraction using rough set as a data mining tool to find the rules between scientific creativity and creativity affective. Deng et. al~\cite{ deng2018innovative} designed a framework for frequent pattern mining using cognitive-based big data analytics. They claim that the discovered patterns assist in the (i) production or extraction of useful knowledge, (ii) identification of behaviour and attitude of users, (iii) comprehension of inference, correlation, inference transmission, and interchange of ideas, and (iv) formulation of suitable decisions and responses.

Singelmann et. al~\cite{ singelmann2020innovators} tried to learn more about the way students solve open-ended questions in an engineering course at the graduate level. They asked all students to submit their own products and accompanying project requirements into an online portal. In this way, they are able to check their progress and then by applying clustering algorithms four clusters appeared. The clusters include Surface Level, Surveyors, Learners, and Innovators. Three taxonomy of learning such as Web's Depth of Knowledge, Bloom's taxonomy, and Cynefin Framework have been used to define those clusters. By observing which students fell into which groups, how they moved among the clusters, and key phrases associated with each cluster, they were able to gain a better understanding of how students confront the creative process. This research gives a clearer picture of the way students develop and find solutions, paving the way for improvements in customized learning, group matching, and even evaluation.

Another study~\cite{lu2010using} also investigated cognitive processes of children by a creativity test using data mining techniques. In this study, they asked 95 fifth grade students to think and apply scientific facts.
A sample board was provided to students to draw as many paths as feasible. The students were also instructed to utilise the fewest possible number of mirrors to hit the target. This variation on the conventional creativity tests allows students to demonstrate their fluency as well as creative thinking. In this experiment, students' drawings were coded and analysed using the association rules mining approach to investigate links between different types of optical paths and to determine students' cognitive abilities in comparison to traditional assessment instruments.

\subsection{Summary}
In this section, we provided background information and discussed the current state of the art in education for detecting and supporting creativity. We first explained educational data and discussed data curation tools and techniques to prepare the data for further analysis. Several known methods are then highlighted and explained such as Educational Knowledge, Educational Data Modeling, OLM, and EDM/LA techniques. In Educational Knowledge, we focused on the concepts of knowledge hierarchy and taxonomy and introduced the traditional methods for creativity assessment.  Next, different data models are explored and discussed how these models, specifically graph data modeling, are used in the education domain. Moreover, OLM as one of the most often used tools for teaching and learning is introduced and investigated for supporting creativity in education. Finally, an overall explanation of data mining in education and LA are provided. The common EDM/LA techniques are disclosed and the important publications in detecting creative thinking patterns are explained to the reader. Table \ref{compare_LR} shows the prominent publications in the literature.

In this study, we leverage to address the inherent flaws in the algorithmic approaches by using a rule-based pattern mining technique as a declarative alternative. The approach has relied on the knowledge of experts in education and focuses on facilitating mining creative thinking patterns from contextualized data and knowledge.

\begin{center}
\begin{table}
\begin{tabular}{|p{2cm}|p{6cm}|p{6cm}|}
\hline
\textbf{Domain} & \textbf{Description} & \textbf{Prominent publications}\\
\hline
Educational Knowledge
&
Using traditional tools and techniques for measuring and detecting creativity.
&
The Torrance Tests of Creativity (TTCT) ~\cite{torrance1972predictive}, gold standard of creativity assessment~\cite{baer2014gold}, self-report measures of creativity~\cite{miller2014self}, and judgment of products~\cite{amabile2018creativity}.
\\
\hline
Educational Data Modeling
&
Organizing the educational data in a way that is suitable for a specific data structure.
&
Hierarchical Data Model ~\cite{teeroovengadum2016measuring, vallerand2002intrinsic}, Network Data Model ~\cite{yu2020construction,sweet2013hierarchical}, Object-oriented Data Model ~\cite{ekwonwune2020multimedia}, Relational Data Model ~\cite{bogdanovic2008development}, and NoSQL Data Model ~\cite{mason2015nosql,franze1999document,martin2007examining}.
\\
\hline
OLM
&
Representing the knowledge, misunderstandings, and challenges of learners.
&
Brainstorming reflection support
~\cite{clayphan2013open}, study of learning to brainstorm
~\cite{clayphan2016wild}, constructing interpretative views of learners ~\cite{ papanikolaou2014constructing}, and supporting metacognition ~\cite{bull2008metacognition}.
\\
\hline
EDM/LA
&
A cycle of data mining and knowledge discovery which is involved with students, instructors or academic authorities, and educational environments to produce educational data and ultimately new knowledge.
&
Frequent pattern mining using cognitive-based big data analytics ~\cite{ deng2018innovative}, rule extraction using rough-set as a data mining tool ~\cite{ yu2010rough}, clustering students for problem-solving skill ~\cite{ singelmann2020innovators}, and associate rule mining to investigate on cognitive processes ~\cite{lu2010using}.
\\
\hline
\end{tabular}
\caption{The prominent publications in the four research areas in the literature: Educational Knowledge , Educational Data Modeling, OLM, AND EDM/LA.}
\label{compare_LR}
\end{table}
\end{center}

\section{Methodology}
This section provides a detailed explanation of the proposed model.
We put the first steps towards formalizing the educational knowledge by constructing a domain-specific Knowledge Base to identify essential concepts, facts, and assumptions in identifying creative patterns~\cite{beheshti2022knowledge}. We introduce a pipeline to contextualize the raw educational data, such as assessments and class activities. Finally, we present a rule-based approach to learn from the Knowledge Base, and facilitate mining creative thinking patterns from contextualized data and knowledge~\cite{shabani2022icreate}. Figure~\ref{pipeline} illustrates the proposed model,
including the
data curation, feature selection,
domain-specific knowledge base, and linking
components.

\subsection{Data Curation}

\begin{figure}[h]
    \centering
    \includegraphics[width=0.9\textwidth]{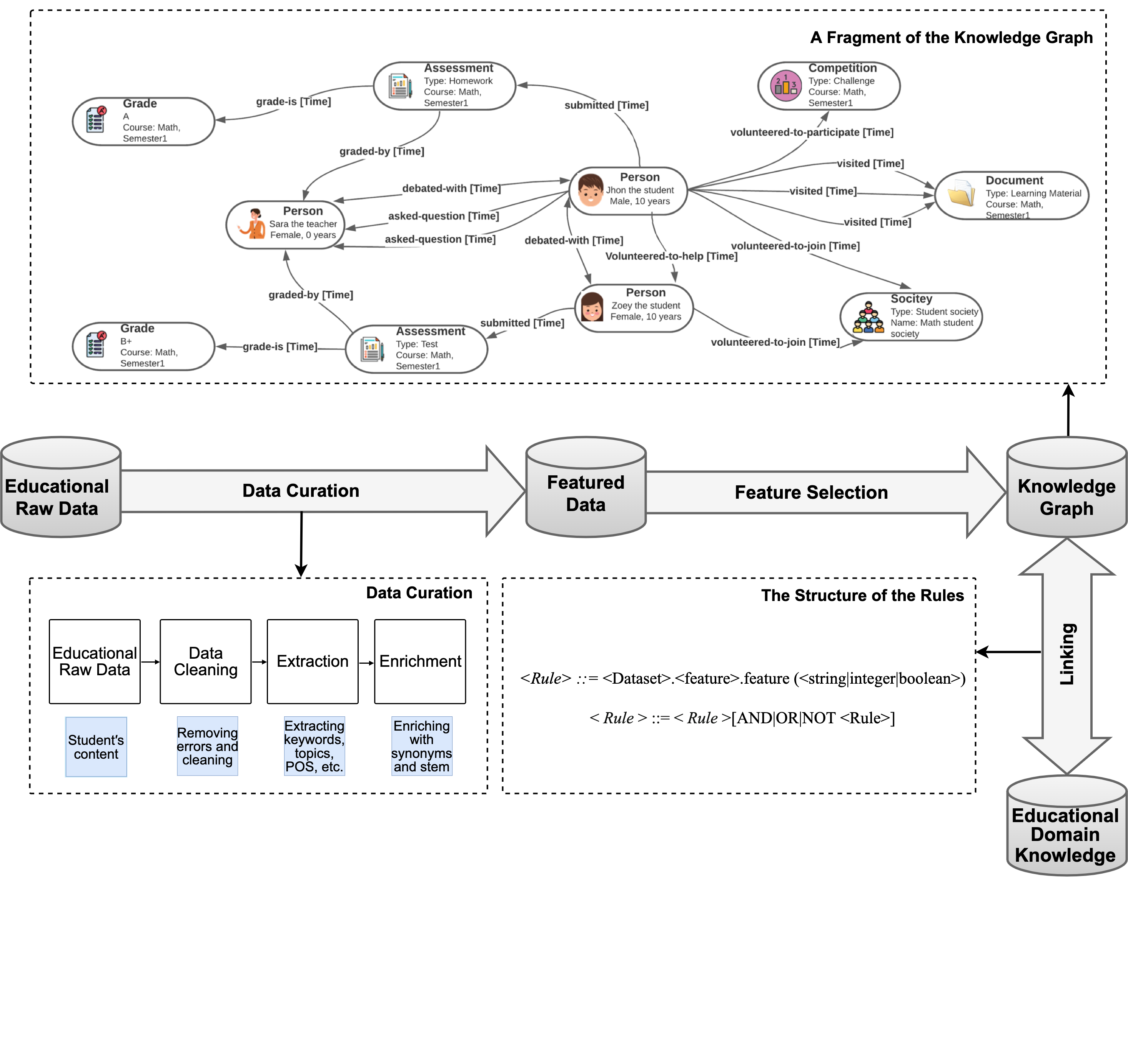}
    \caption{An overview of the proposed model.}
    \label{pipeline}
\end{figure}

In this step, we leverage the state of the art in data curation~\cite{beheshti2018corekg, beheshti2019datasynapse}, to turn the raw educational data into contextualized data and knowledge. As it is shown in Figure~\ref{pipeline}, the curation pipeline includes cleaning, extraction, and enriching phases.
For example, considering a student's assessment as a raw data, the curation pipeline will:
(i)~clean raw data, by identifying and correcting inaccurate or corrupt information, such as missing, erroneous, or irrelevant data, and then correcting them;
(ii)~extract features (such as keywords, entities, named entities, topics, POS) from the contents; and
(iii)~enrich the extracted features with synonyms, stems, and similar keywords and entities to make it ready for feature selection and further analysis.

\subsection{Feature Selection}

Reducing the dimensionality of data has multiple benefits, which make the dataset less complex and thus easier to work.
Models are also less prone to overfit on a dataset with fewer dimensions~\cite{khalid2014survey}. The simplest technique to minimize dimensionality is to choose only the key features from a huge dataset. The readability and performance of the model are heavily influenced by the variables chosen from the student dataset's list of variables. This would be accomplished by using a feature selection approach on a dataset.
Since it is critical to find which features significantly influence the pattern discovery process, the features will be selected from the enriched data. We aim to employ several types of features in education,
including:
%
(i)~Demographic data, that contain information about the characteristics of students such as age, gender, attendance, economic status, enrollment, dropout rates, language proficiency, disabilities, student mobility, and behavior problems;
(ii)~Achievement data, that contain information about student achievement and learning outcomes such as test scores, homework, class-based scores, and rubric-based portfolios; and
(iii)~Program data, that contain information about a school's programs such as its academic programs, teacher training and experience, professional development, and extracurricular programming.

\subsection{Domain-specific Knowledge Base}

The next important step here is to imitate the knowledge of experts in education into an Educational Knowledge Base, which provides a rich structure of relevant entitles, semantics, and relationships among them. A Domain Knowledge (DK)
is a taxonomy of concepts, sub-concepts, and their relationships.
In section~4, we focus on a motivating scenario in education domain and present the construction of such Domain Knowledge.

\subsection{Educational Knowledge Graph}

As introduced in Section~\ref{knowledge_graph_RW}
, a Knowledge Graph
is a set of interconnected entity descriptions (real-world objects or concepts) that combines characteristics of other data management systems like graph databases. It is basically a knowledge base that has been made machine-readable using logically coherent, connected graphs that form an interconnected set of facts.
RDF-based knowledge graphs~\cite{hammoud2015dream,zhang2021fraudre} are the best foundation for data integration and unification in this context. RDF or Resource Description Framework is made to model schemaless databases for the Semantic Web flexibly. Its structure is based on triples, each of which is made up of three entities that formalize semantic data in the form of the subject $\rightarrow$ predicate $\rightarrow$ object. A triple also referred to as statements' or RDF statements', signifies a relationship between subject and object predicate captures. As a result, a directed graph, with nodes representing subjects and objects, and edges representing predicates, can be used to describe a set of triples~\cite{triplestore,batarfi2015large}.
For example, the RDF statement Sara submitted homework' can be defined in a triplestore and explains the relationship between the sentence's subject, "Sara" and the object, "homework" The predicate "submitted" expresses the relationship between the subject and the object.
As a graph database, Triplestore stores data as a network of objects connected by materialised connections. As a result, RDF triplestores are the best option for organizing heavily interrelated data.

In this study, based on the aforementioned facts, to extract insight from the
contextualized data,
%
we construct an RDF-based Graph to represent the knowledge hidden in the contextulaized data (i.e., the result of the proposed data curation pipeline).
We call this graph the Educational Knowledge Graph (eKG).
Let $R = (Ent, Rel)$ be an RDF graph where \textit{Ent} is a set of entities and Rel is a set of relationship labels. $Let G = (V, E)$ be an Entity-Relationship (ER) attributed graph where \textit{V} is a set of nodes and E is a set of ordered pairs called edges. An \textit{ER} graph $G_{eKG} = (V_G, f_{Ent}, E_{Rel})$ where \textit{V} is a set of nodes, $f_{Ent}: V_G \rightarrow E_{Rel} $ is an injective function, and $E_{Rel}$ is a set of labeled edges. $G_{eKG}$ with n number of entities is defined as $G_{eKG} \in R $, $V_G = V$, and $E_{R} \in Rel $.
The graph is viewed as a directed graph since it is always possible to interpret the direction of a relationship between two entities in the opposite direction.

The eKG graph is composed of entities and relationships among them.
An entity is a real-world object that has unique existence and can be distinguished from other objects by physical (e.g., a teacher, a student) or conceptual (e.g., a course, a task) existence. They are defined by a set of attributes, e.g., name, age, gender.
A relationship can be defined as a directed link between two or more entities that are defined by a predicate based on the attributes of the entities. 
We model the featured data as a graph of typed nodes and edges (relationships). The upper part of the Figure~\ref{pipeline} depicts a small fragment of a knowledge graph showing possible relationships between objects in an educational setting (A comprehensive example of entitles and their relationships are presented in Section 4.1.).
The entities and relationships create different type of triples. Few examples of the triples depicted in the graph are described as follows:
    \begin{itemize}
        \item Assessment $\xrightarrow{\mathit{graded-by [Timestamp]}}$ Person: States that an assessment (e.g., a homework) is graded by a person (e.g., a teacher).
        \item Person $\xleftrightarrow{\mathit{debated-with [Timestamp]}}$ Person: States that two person debated over a topic (e.g, a teacher and a student debated with each other).
        \item Assessment $\xrightarrow{\mathit{grade-is [Timestamp]}}$ Grade: States the grade of an assessment.
        \item Person $\xrightarrow{\mathit{submitted [Timestamp]}}$ Assessment: States that a person (e.g., a student) submitted an assessment.
        \item Person $\xrightarrow{\mathit{volunteered-to-join [Timestamp]}}$ Society: States the a person (e.g., a student) is volunteered to join a society (e.g., a student society).
    \end{itemize}

The Educational Knowledge Graph will make it easier to find, evaluate, and communicate essential patterns in educational data, allowing us to investigate the possibilities for better understanding and evaluating student behavior and performance. 

\subsection{Rule-based Insight Discovery}

The final step is to derive relevant insights from the data in order to reach a consensus. Insight discovery is the process of extracting evidence from data to help analysts get a precise and in-depth insight into a specific analytic purpose. To this end, we perform user-guided insight discovery~\cite{beheshti2019datasynapse} to facilitate the insight discovery process and link extracted featured items to the entities in the domain-specific knowledge by using a simple rule language.
%
Many of the flaws inherent in algorithmic approaches can be addressed using rule-based techniques as a declarative alternative. Algorithms are suitable for tackling particular tasks, but datasets are huge and changing all the time in reality. There are various advantages of using rule-based techniques over algorithmic designs such as being easier to produce, faster for correcting errors, and wider coverage of cases \cite{ beheshti2019datasynapse, gc2015big}. Generally, rules can be defined as follows:
\begin{equation} \label{eq:1}
<Rule> ::= <Dataset>.<feature>.feature (<string|integer|boolean>)
\end{equation}

Rules can become more complicated by using conjunction or dis-junction of additional rules. 
\begin{equation} \label{eq:2}
<Rule> ::= <Rule> [AND|OR|NOT <Rule>]
\end{equation}

Ultimately, an IF THEN rule can be used to assign a tag and link the extracted featured items to the domain knowledge entities.

\begin{equation} \label{eq:3}
  <Rule> =
    \begin{cases}
       \text{IF <Rule1> [AND|OR|NOT <Rule2>]} \\
       \text{THEN}\\
       \text{TAG[<Dataset>, <feature>.feature(True)]}
    \end{cases}
\end{equation}

Example: To identify a creativity pattern (e.g., strong memory) in a student two rules can be defined and connected to flag a tag as follows:

\begin{equation} \label{eq:4}
  <Rule> =
    \begin{cases}
       \text{<Rule1> = Student(ID).Assessment(Type, Calender).assessmentScore.Graded[integer]}\\
       \text{<Rule2> = Student(ID).classActivity.visited.learningMaterials[integer]}\\
       \text{IF <Rule1> AND <Rule2>} \\
       \text{THEN}\\
       \text{TAG[Student, Creativity.Cognitive.strongMemory(True)]}
    \end{cases}
\end{equation}

It should be noted that to understand essential features in the rules, we have interviewed education experts and also conducted surveys to mimic their knowledge.  In future work, we plan to use Human-in-the-loop (crowdsourcing) approaches to create the rules and minimize the manipulation of the rules.

\textbf{Querying RDF Graph.}
To enable rule-based querying of the RDF stores in the Educational Knowledge Graph, we intend to use the SPARQL query language for analysing and organizing the extracted-enriched data and linked features. SPARQL is a SQL-like edge-based query language for RDFs that will be used on top of a graph to search through the nodes and edges \cite{farouk2012rule}.
A SPARQL query \textit{Q} is defined as a tuple $Q = (AE, DS, OR)$. It is built around an algebra expression \textit{AE} that evaluates an RDF graph in a dataset \textit{DS}. \textit{AE} is made up of several graph patterns that can include solution modifiers, including LIMIT, ORDER BY, DISTINCT, PROJECTION. The outcomes of the matching process are handled using the operation result OR (e.g., CONSTRUCT, SELECT, ASK, FILTER, DESCRIBE).
%
A basic SPARQL query has the following format:

\footnotesize
\begin{verbatim}
 select ?variable1 ?variable2 ...
 where { pattern1. pattern2. ... }
\end{verbatim}
\normalsize

Each pattern has three parts: subject, predicate, and object, which can all be variables or literals. The known literals are specified in the query, while the unknowns are left as variables. To answer a query we should locate all potential variable bindings that meet the given patterns.
The '@' indication is used to identify attribute edges from the relationship edges between graph nodes. Example 5 presents a sample graph-level query.
The triple pattern t is the most straightforward graph pattern specified in SPARQL. The subject, predicate, and object variables can be used in a triple pattern analogous to an RDF triple. Triple patterns, like triples, can be represented as directed graphs. Hence, a SPARQL query is frequently cited as a graph pattern matching problem~\cite{huang2011scalable}.

\begin{equation}
  t \in TP = (RDF-T \cup V ) \times (I \cup V ) \times (RDF-T \cup V )
\end{equation}

\begin{figure}[t]
    \centering
    \includegraphics[width=0.7\textwidth]{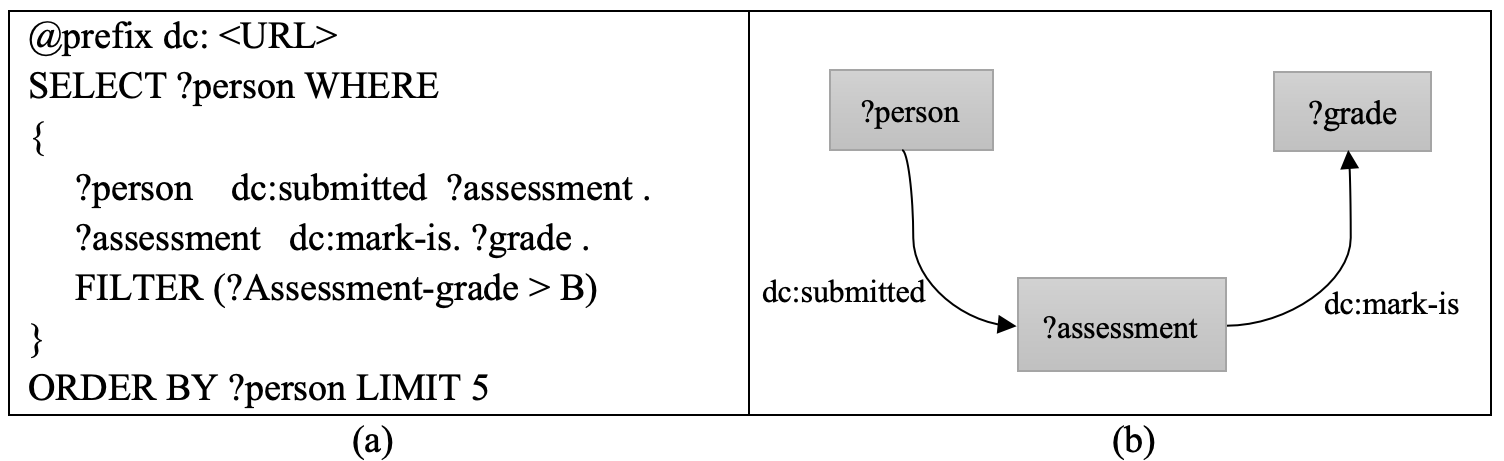}
    \caption{(a) A SPARQL query with one FBGP that searches for the name of the students that submitted an assessment and received grades greater than B; and (b) Graph pattern showing the pattern that will be extracted from the dc RDF.}
    \label{sparql1}
\end{figure}

where RDF-T denotes the set of RDF terms, I denotes a set of IRIs, and V denotes a set of variables.
A basic graph pattern $BGP = \{ t_1 ... t_n \}$ is defined as a set of triple patterns with $t_1 ... t_n \in TP$. If all of the enclosed triple patterns match, it matches a subgraph. Value constraints (FILTER) and other graph patterns can be combined with basic graph patterns. Basic graph patterns and value constraints are evaluated in a non-ordered manner. This indicates that a structure consisting of two basic graph patterns BGP1 and BGP2 controlled by constraint C, could be converted into a single equivalent basic graph pattern followed by a constraint. Filtered basic graph patterns referred to as FBGP are basic graph patterns that have been modified by one or more constraints~\cite{w3}.

\textbf{Example 1}:
Figure \ref{sparql1}
shows a SPARQL query with one FBGP that search for the name of the students that submitted an assessment and received grades greater than B. The results are ordered by name with a maximum of five results.
In this example, ?x plays as subject, dc:name as predicate, and ?name as object in a triple. The question mark comes before variables and literals come between quotations in the queries. FILTER eliminates solutions that do not cause an expression to evaluate to be true. And ORDER BY sorting the results. Finally, the results will be shown in XML or JSON format.

\begin{figure}[t]
    \centering
    \includegraphics[width=0.7\textwidth]{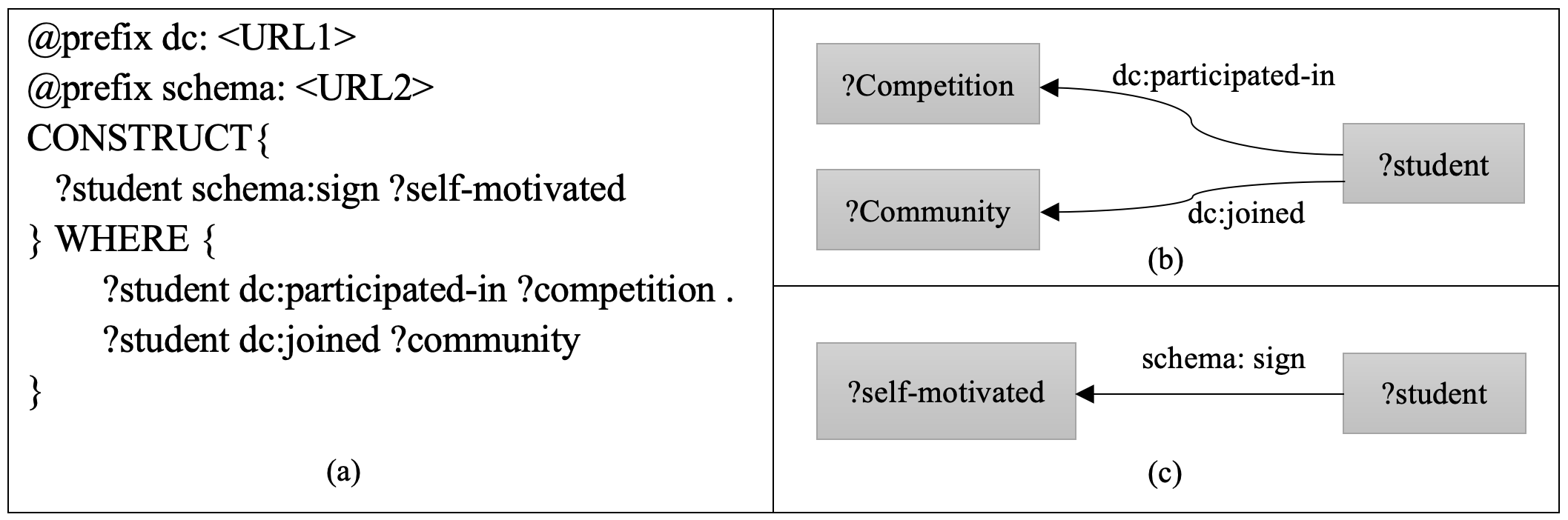}
    \caption{(a) A SPARQL query with a CONSTRUCT query that makes a new RDF statement; (b) The initial graph; and (c) The constructed graph out of db.}
    \label{sparql2}
\end{figure}

\textbf{Example 2}:
Figure~\ref{sparql2}
shows a more complex SPARQL query using CONSTRUCT statement that returns a new RDF graph out of the different graph datasets. In this example, we are looking for a pattern of motivation as a student who is involved in participating in competitions and social communities.

\section{Experiments and Evaluation}

In this section, we demonstrate and evaluate the outcomes of the
proposed method (section~3)
and examine how this method
can be
used to mine patterns of creativity in individuals. We first discuss a motivating educational scenario to clarify our approach toward mining creativity patterns. Next, we provide detailed information on the input data and experimental setting, present the experimental results, and finally explain how the results are evaluated.

\subsection{Motivating Scenario}

Creativity is seen as a fundamental quality that children today must possess in order to excel in school, and the career~\cite{amabile1998kill}. From big-picture planning to rigorous organising, this 21st-century talent empowers students to express their inherent strengths. They learn about their creativity as well as how to use it in a healthy and effective manner.

As a motivating scenario, we focus on detecting creativity patterns in students to help teachers find creative students in the classroom. They can also find struggling students and plan for fostering creativity in those. For example, the plan would be to detect motivated, creative students and get help from them to lead groups of students and share their knowledge with others. This spreads creativity; when one student comes up with a unique and interesting approach to solve the problem, after sharing the solution with other students, it is possible that the next student will be encouraged to do something similar. 

To facilitate discovering patterns of creativity in students, we propose these three important steps as follows:

\textbf{Use Case 1: Imitating the Knowledge of Experts in Education.}
The initial phase in the pattern discovery is building a KB by imitating the knowledge of education experts who have the best knowledge in education and students' skills. The KB consists of a set of concepts organized into an educational taxonomy, instances for each concept, and relationships among them. We explain the techniques we used to construct the KB domain knowledge.

\begin{sidewaysfigure}
    \includegraphics[width=\columnwidth]{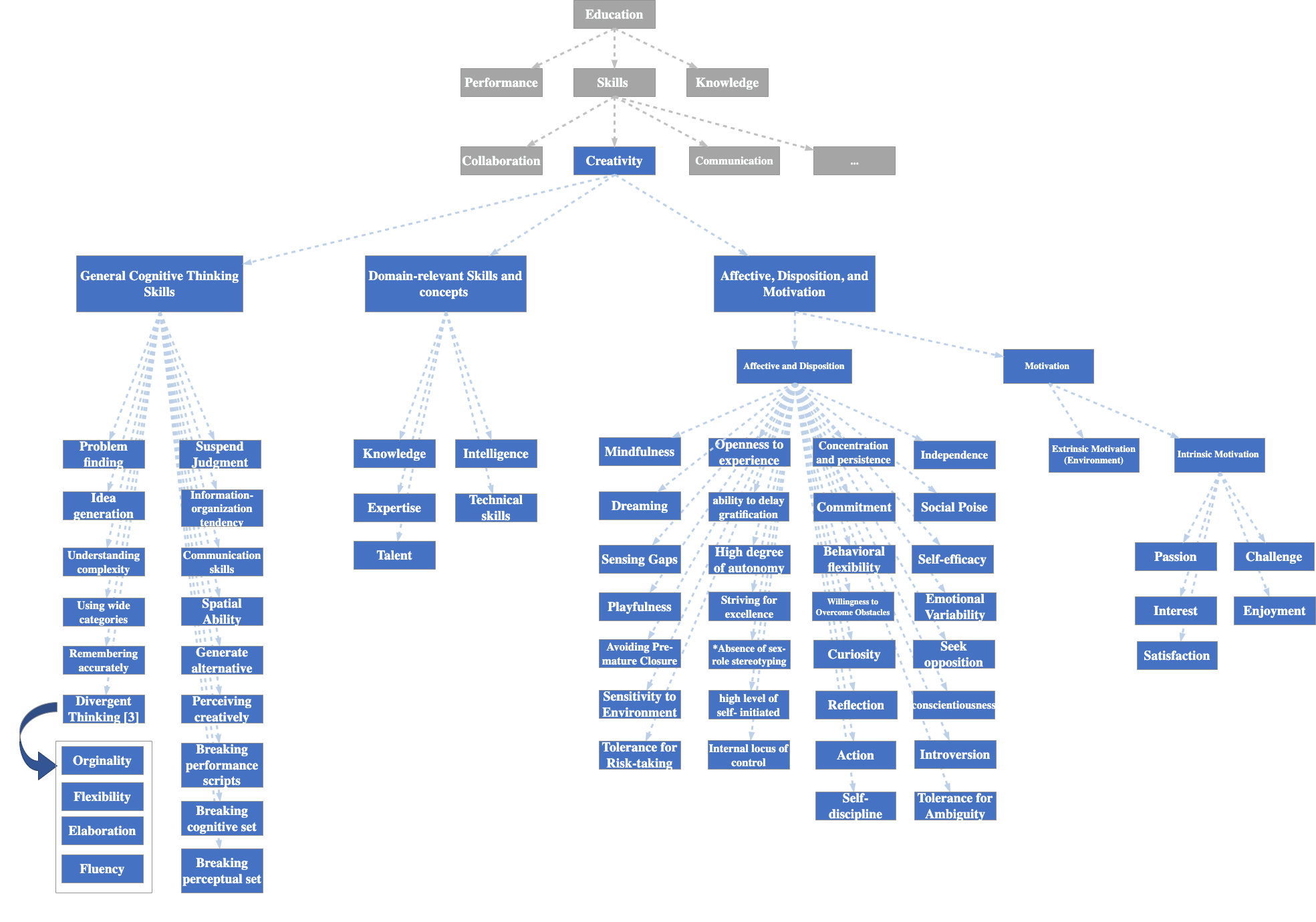}
    \caption{A fragment of the Educational Knowledge Base focusing on creativity concepts.}
    \label{ourtaxonomy}
\end{sidewaysfigure}

We spent several months studying major articles and books in education and cognitive research in order to compile a list of the significant notions of creativity and their sub-concepts. We based our work on the publication series of Teresa Amabile \cite{amabile2011componential}, and Bloom's taxonomies \cite{bloom1956taxonomy} in combination with other major works exploring the instances of creativity \cite{guilford1973characteristics,james2001personality,sternberg1999creativity}. We finally formalized them with the help of an expert in education and created a creativity taxonomy shown in Figure~\ref{ourtaxonomy}. The main concepts of the taxonomy are identified as follows:

\begin{itemize}
    \item General Cognitive Thinking Skills: The mental processes involved in gaining knowledge and comprehension. These cognitive thinking processes include idea-generating, remembering, using wide categories, and problem finding skills;
    \item Domain-relevant Skills and Concepts: The amount to which a person's product or reaction will outperform past responses in the domain is determined by his or her usage of creativity-relevant abilities. It includes expertise, knowledge, technical skills, intelligence, and talent in the particular domain;
    \item Affective, Disposition, and Motivation: Affective and Disposition include the ways in which students deal with external and internal phenomena emotionally such as self-efficacy, independence, curiosity, and commitment. Furthermore, motivation encompasses both intrinsic and extrinsic factors such as passion, challenge, interest, enjoyment, and satisfaction.
\end{itemize}

\textbf{Use Case 2: Educational Knowledge Graph.}
As is explained in Section 3.4, an RDF graph will be built from the extracted features. The graph consists of two components, entity and relationship, which together create triples of instances. The entities are real-world objects with a distinctive physical (e.g., a teacher, a student) and conceptual (e.g., a course, a task) identity that distinguishes them from other objects. They are defined by a set of attributes, e.g., name, age, gender, nationality, and student ID. In our scenario, entities have name types as School, Person, Course, Training Course, Competition, Assessment, Question Block, Tool, Document, Task, Student Society, Time, Typography Error, Assessment Score, Word-count, Keyword, and Sentiment Analysis.

Relationships that are directed links between entities can be defined by a predicate based on the attributes of the entities. Different types of triples that are shown in Figure \ref{eKG} are described as follows:

\begin{figure}[h]
    \centering
    \includegraphics[width=1\textwidth]{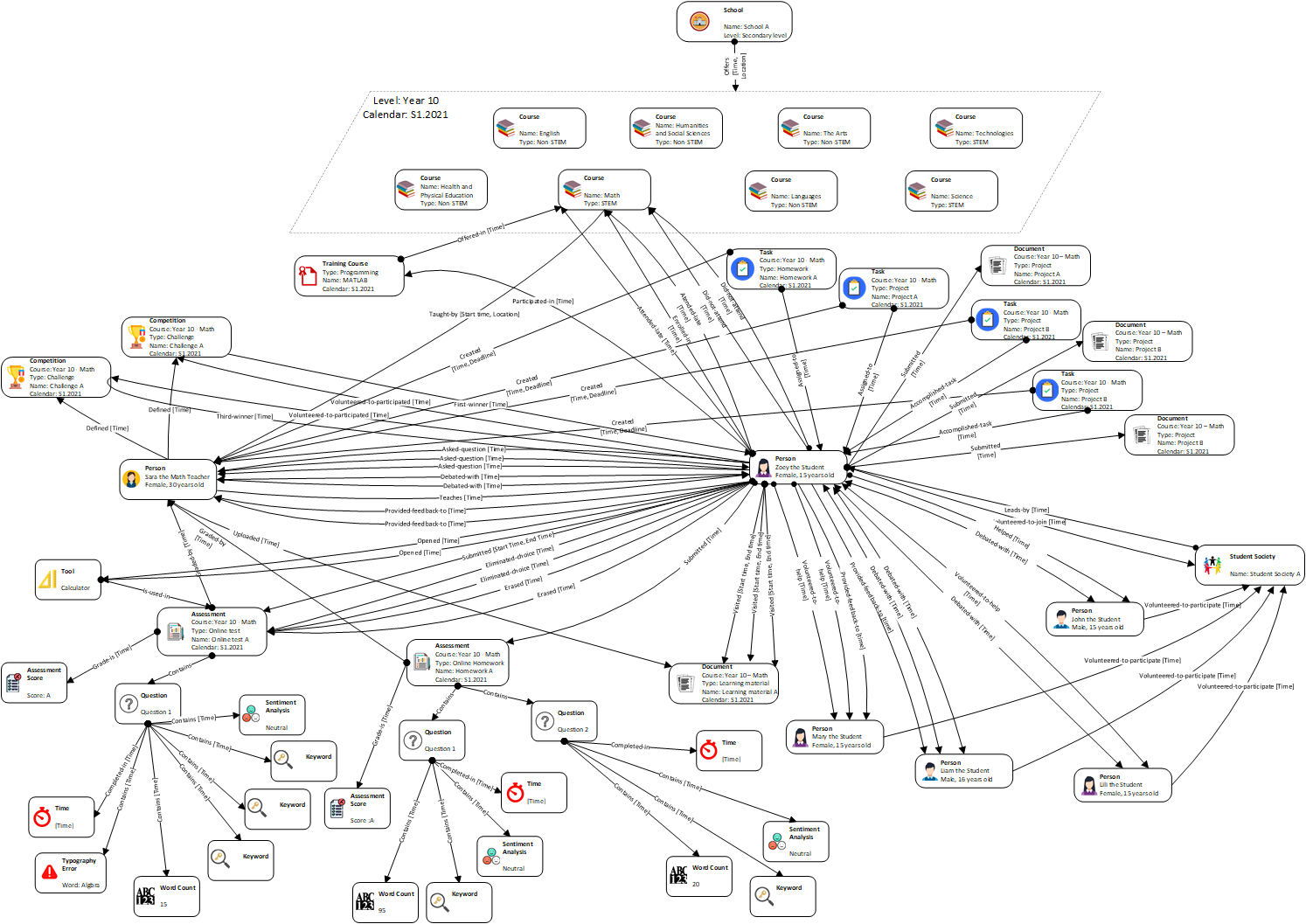}
    \caption{A canonical graph of the possible relationship between objects in an educational setting.}
    \label{eKG}
\end{figure}

    \begin{itemize}
        \item Course $\xrightarrow{\mathit{taught-by [Timestamp]}}$ Person: States that a course (e.g., Math) is taught by a person (e.g., a teacher).
        \item Person $\xrightarrow{\mathit{submitted [Timestamp]}}$ Assignment: States that a person (e.g., a student) submitted his/her assignment (e.g., a homework).
        \item Person $\xrightarrow{\mathit{debated-with [Timestamp]}}$ Person: States that two persons debated.
        \item Person $\xrightarrow{\mathit{volunteered-to-help [Timestamp]}}$ Person: States that a person (e.g., a student) volunteered to help another person.
        \item Assessment $\xrightarrow{\mathit{contains}}$ QuestionBlock: States that an assessment (e.g., an online math test) contains a question block (e.g., fill-in question, multiple choice).
        \item Assessment $\xrightarrow{\mathit{graded-by [Timestamp]}}$ person: States that an assessment (e.g., an online Math test) is graded by a person (e.g., a teacher).
         \item QuestionBlock $\xrightarrow{\mathit{contains}}$ Keyword: States that a question block of an assignment (e.g., a homework) contains desired keywords relevant to the context.
        \item Person $\xrightarrow{\mathit{volunteered-to-join [Timestamp]}}$ StudentCommunity: States that a person (e.g., a student) volunteered to join a student community (e.g., Math community of students)
        \item QuestionBlock $\xrightarrow{\mathit{completed-in [Timestamp]}}$ Time: States that a question block of an assessment is completed in a specific amount of time (e.g., 2 min or 120 sec).
        \item Person $\xrightarrow{\mathit{volunteered-to-participate [Timestamp]}}$ TrainningCourse: States that a person (e.g., a student) volunteered to participate in a training course (e.g., MATLAB programming).
    \end{itemize}

\textbf{Use Case 3: Linking the Knowledge base to the Knowledge Graph}
We perform a user-guided insight discovery task to analyze the knowledge graph and identify creativity patterns related to each student node. In this step, we link the extracted features to the concepts of creativity in the eKB. To this end, we got help from educational experts and defined a set of rules for each pattern to guide the process of finding related subgraphs. For example, four extracted subgraphs and related rules are depicted in Table \ref{rule_examples}.

\begin{table}
  \caption{Four example of creativity rules and their related patterns.}
  \label{rule_examples}
  \includegraphics[width=\linewidth]{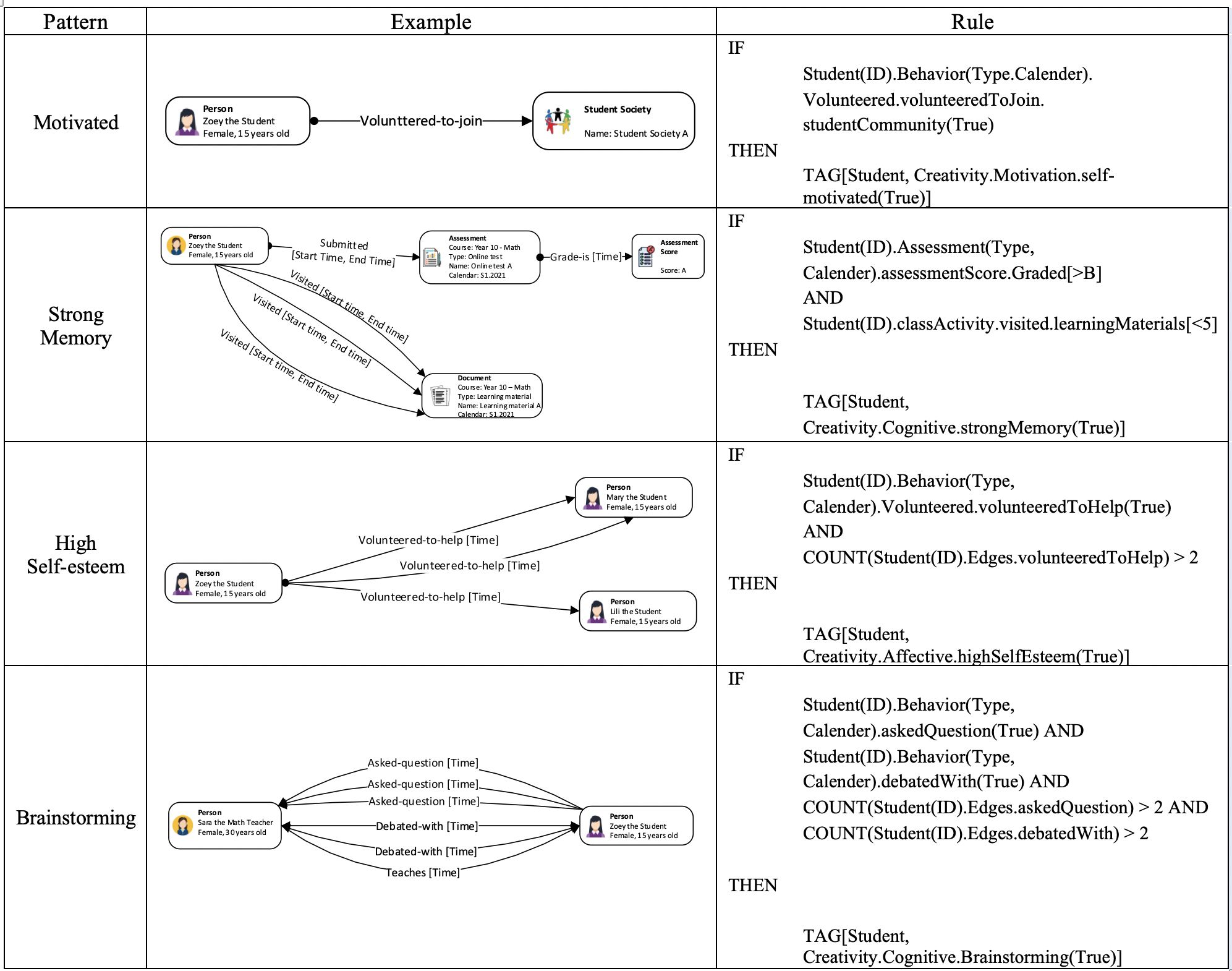}
\end{table}

\subsection{Dataset}
We used a Kaggle public educational dataset \footnote{https://www.kaggle.com/aljarah/xAPI-Edu-Data} which is gathered via the experience API, a learner activity tracker tool (xAPI). The xAPI is a part of the Training and Learning Architecture (TLA) that allows tracking the learners' progress and activities, such as reading, watching, and viewing learning materials.

The dataset contains information about 480 students with 16 features which are divided into three groups: (1) Academic background features such as educational level, section, and stage. (2) Behavioral features include the number of raised hands in class, involved discussions, and opening resources. (3) Demographic features such as nationality, gender, and age.

The dataset is made from 175 females and 305 males. The students are from various countries (e.g., USA, Morocco, Iran, and Venezuela) with different backgrounds and level of education. The data was gathered over the course of two academic semesters and contains a school attendance feature, in which students are divided into two groups based on the number of days they are absent (under seven days and above seven days). This dataset also includes features related to student's behavior in the class (e.g., raising a hand, discussion, and visiting resources) in different courses (e.g., IT, Math, English) during a semester. Based on their overall mark, students are divided into three numerical ranges: Low-Level (marks between 0 to 69), Middle-Level (marks between 70 to 89), and High-Level (marks between 90 to 100).

\subsection{Experimental Setting}
The experiments are performed on two platforms: Google Colab and GraphDB graph database. We used Python 3.10.0, Pandas 1.3.4, Neworkx 2.6.2, and rdflib 6.0.1 for building the RDF Knowledge Graph. The .ttl (Turtle RDF graph) version of the graph model is then imported to the  GraphDB environment to perform SPARQL queries.

\subsection{Experimental Results}
\subsubsection{Preprocessing and Feature Selection}
After importing the mentioned dataset, we performed some preprocessing tasks such as data cleaning, removing noisy inputs to avoid errors, and renaming the column names and instances to make the graph and queries more readable and clearer. Then we selected the features that are most relevant to our approach. Table \ref{dfdataset} shows the first five rows of the selected features. Since the data is anonymized, the students are identified by numbers and indexes.

\begin{table}
  \caption{First five rows of the dataset after feature selection.}
  \label{dfdataset}
  \includegraphics[width=\linewidth]{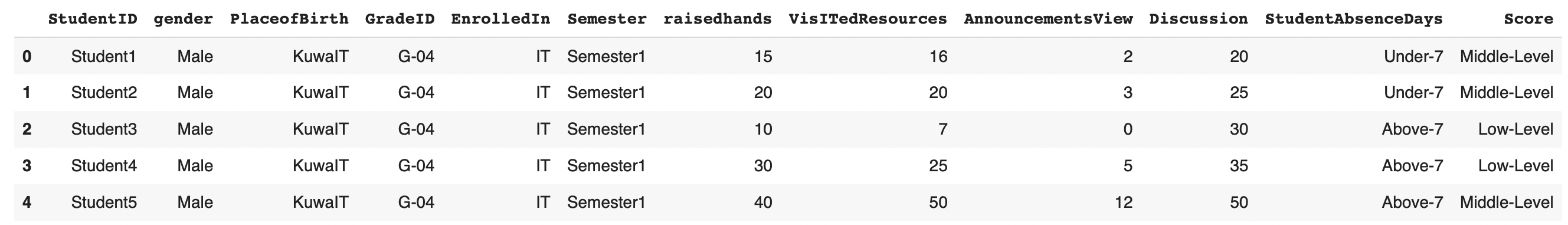}
\end{table}

\subsubsection{Building the Knowledge Graph}
\begin{figure}[htp]
    \centering
    \includegraphics[width=0.8\textwidth]{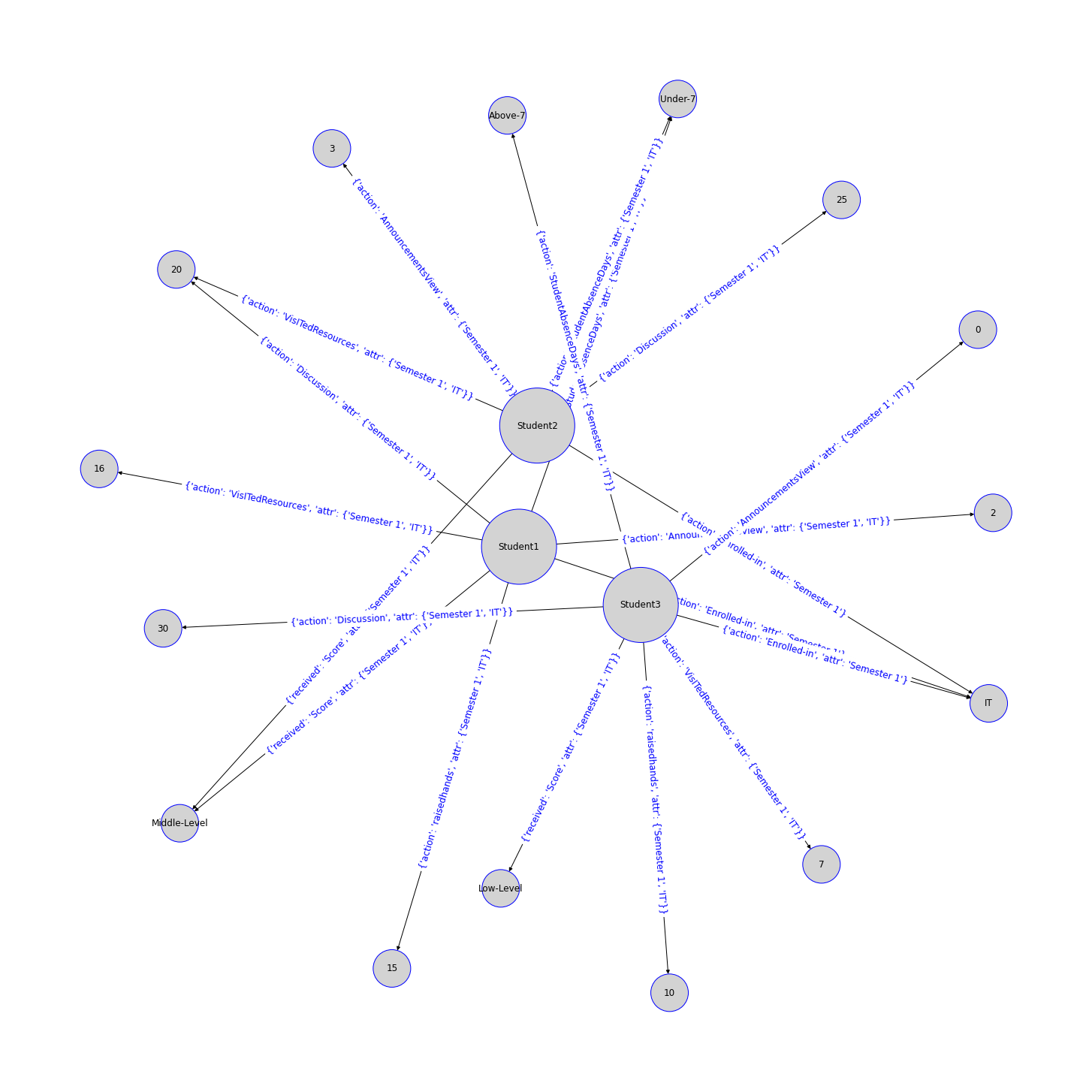}
    \caption{Converted data from rows of data to triples (object, predicate, object) for Student1, Student2, and Student3.}
    \label{three_students}
\end{figure}

The next step is to build an RDF graph out of the selected features. For clarity, we separated a subgraph from the completed graph and showed it in Figure \ref{three_students}. The figure depicts the triples related to three students with their relevant attributes (e.g., Gender, Place of birth, and Grade). The entities and relationships of the rdf subgraph are listed as follows (for Student1):
\\
@prefix ns1: <http://www.example.org/> . \\
ns1:Student1 ns1:AnnouncementsView 2 ;\\
ns1:Student1 ns1:Discussion 20 ; \\
ns1:Student1 ns1:EnrolledIn ns1:IT ;\\
ns1:Student1 ns1:Score ns1:Middle-Level ;\\
ns1:Student1 ns1:Semester ns1:Semester1 ;\\
ns1:Student1 ns1:StudentAbsenceDays ns1:Under-7 ;\\
ns1:Student1 ns1:VisITedResources 16 ;\\
ns1:Student1 ns1:raisedhands 15 .\\

The completed graph then will be saved in .ttl format to be imported into a graph database. To query this huge graph, we make use of SPARQL queries to organize the data and extracted features.

\subsubsection{Linking the Graph to the Knowledge Base}
To link the constructed RDF graph to the Educational Knowledge Base (explained in Section 4.1), we define a set of rules to be implemented on top of the RDF graph repository. As a result, students that follow the rules will be shown in the result of the queries.

For instance, three rules are applied across the graph, including those about brainstorming, having a great memory, and a lack of motivation. Brainstorming is a concept related to the "Using wide range of categories" concept under "General cognitive thinking skill". A strong memory is also a concept related to the "Strong memory" concept under "General cognitive thinking skill". Furthermore, Motivation is the central concept of creativity that can be found in the creativity hierarchy under the name of "Intrinsic Motivation".

The related SPARQL queries with their node results are shown in Table \ref{sparql_query}.
The queries are specifically looking for creative students enrolled in Math class in Middle School during either semester 1 or 2. As you can see, just four students showed the pattern of "Strong Memory" in their behaviors. This rule involves the students' marks and the number of visiting resources. Students who gained high marks in their assessments and visited resources lower than 20 times (in comparison with others) during a semester are considered to have strong memory. It is specifically important in Math, Science, and IT courses.

Ten students showed the pattern of "brainstorming" in their behaviors. This rule involves the number of asking questions and discussions in groups. Also, two students did not show the pattern of "Motivation" in their behaviors. This rule involves attending classes and viewing announcements to participate in different events.

\begin{center}
\begin{table} [h]
\begin{tabular}{|c|}
\hline
SPARQL Query and Results\\
\hline
Strong Memory\\
\includegraphics[scale = 0.9]{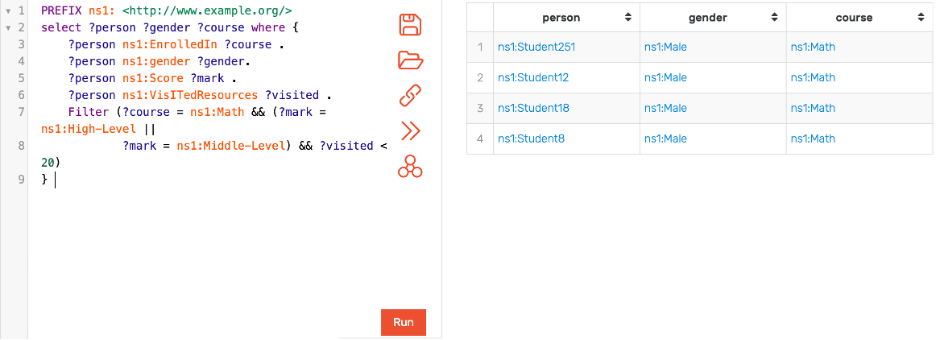}\\
\hline
Brainstorming\\
\includegraphics[scale = 0.9]{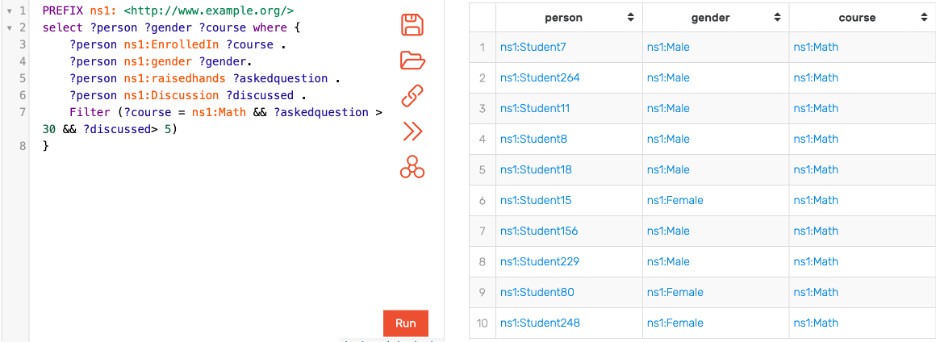}\\
\hline
Not Motivated\\
\includegraphics[scale = 0.9]{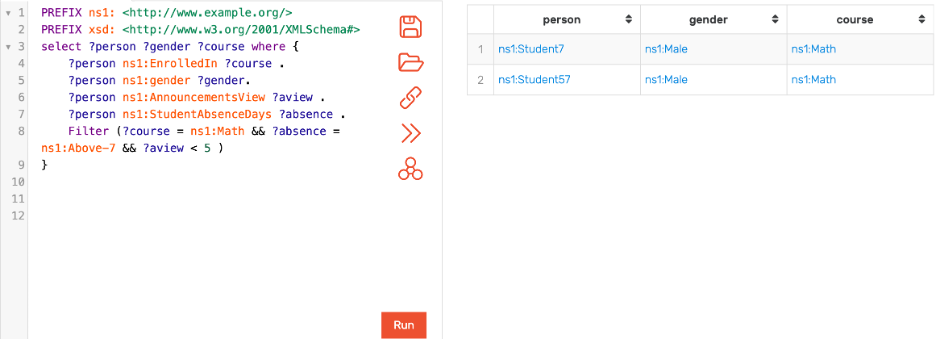}\\
\hline
\end{tabular}
\caption{The SPARQL queries and their results for finding students with three creativity patterns: Strong memory, Brainstorming, and Motivation patterns.}
\label{sparql_query}
\end{table}
\end{center}

\subsection{Evaluation}
To evaluate the correctness of the model, we carried out a user study. In this study, we tried to validate the following hypotheses.

\begin{itemize}
    \item H1: The components of the KB are relevant to creativity.
    \item H2: The designed rules for pattern mining are useful and relevant to creativity patterns.
    \item H3: The query results are reasonable and show a successful link between the KB and educational data.
\end{itemize}

\subsubsection{Experiment Setup}
The experiment was done in a controlled environment to examine our approach. Participants were mostly chosen among academics and students doing research at Data Analytics Research Lab \footnote{https://data-science-group.github.io/} with different backgrounds. Hence, from 10 participants some had strong knowledge in education and cognitive science, some had computing domain expertise, and others had both.
We first instructed participants about the educational motivating scenario and demonstrated the functionality of the model through a presentation with the following order:

\begin{enumerate}
    \item Imitating the Knowledge of Educational Experts: We first underlined the importance of building the KB and how this helped us to link related information in the educational data and components of creativity in education.
    \item Data Contextualization: We explained how using existing data curation techniques helped us create enriched-contextualized data and knowledge.
    \item Linking Data and Finding Patterns: We presented a fragment of the data in a visualized graph-based format to be easily understood. We also demonstrated the results and implemented the defined rules for each creativity pattern.
\end{enumerate}

\subsubsection{Questionnaire}

We prepared a questionnaire and shared it with the participants to examine the study's hypotheses. The questionnaire consisted of four parts having multiple choice questions and participants were instructed to only choose one option based on their interpretation. With each of the participants, we asked them to rate the relevancy of the hypotheses using a Likert scale system which uses a five-point scale to allow the participant express their opinion (5: Strongly relevant, 4: Relevant, 3: Neutral, 2: Weakly relevant, 1: irrelevant).

The first part of the questionnaire was about participants' demographic and background information and the rest was designed to evaluate H1, H2, and H3 hypotheses. The questionnaire was designed by Google Forms and
a few snapshots can be found in Table \ref{Questions} for evaluating the "using wide range of categories" creativity pattern. In this survey, we provided definitions and hints for the participants and guided them through the whole process.

\begin{center}
    \begin{table} [h]
    \begin{tabular}{c c}
    \includegraphics[scale = 0.4]{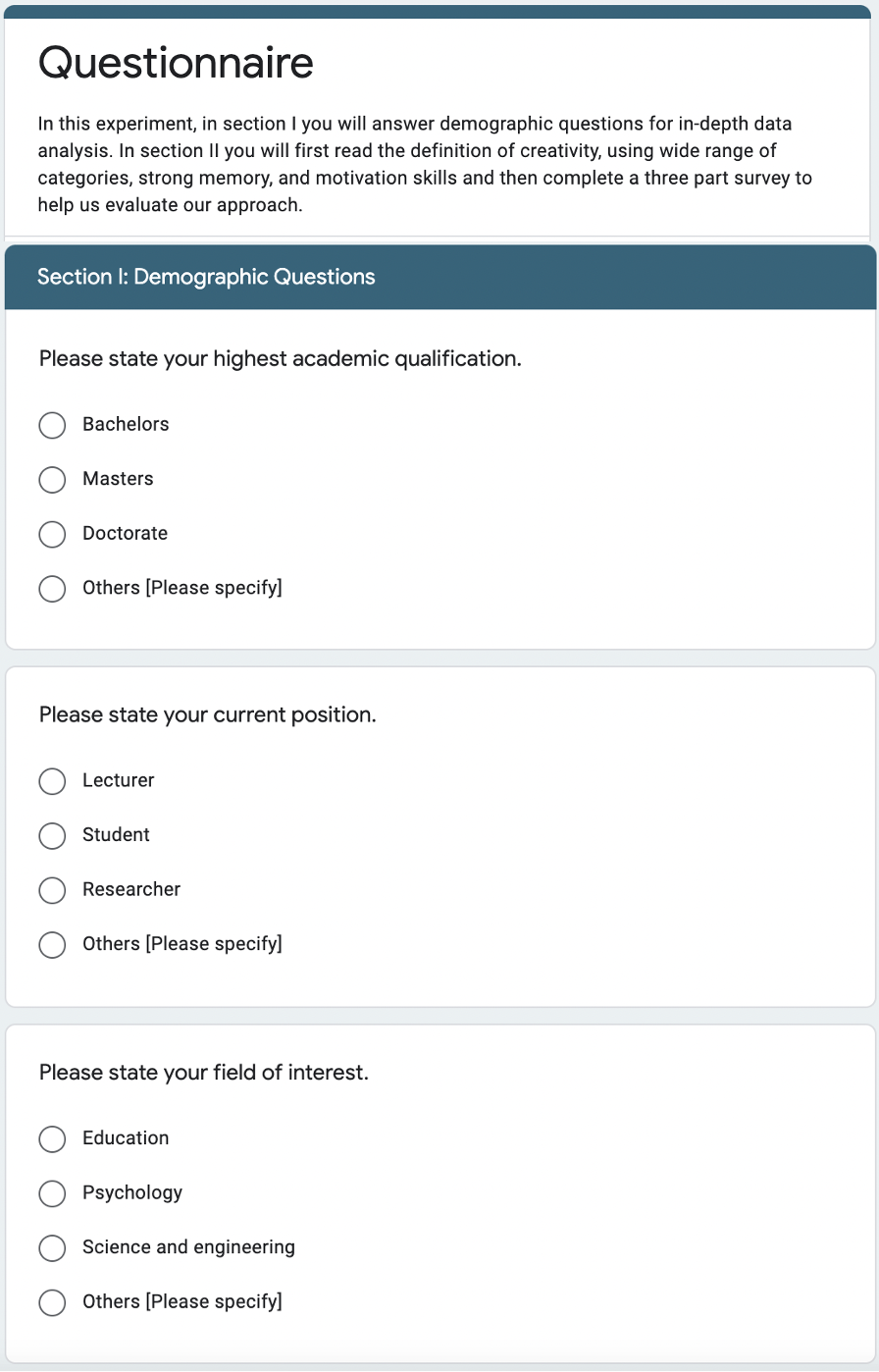} & \includegraphics[scale = 0.4]{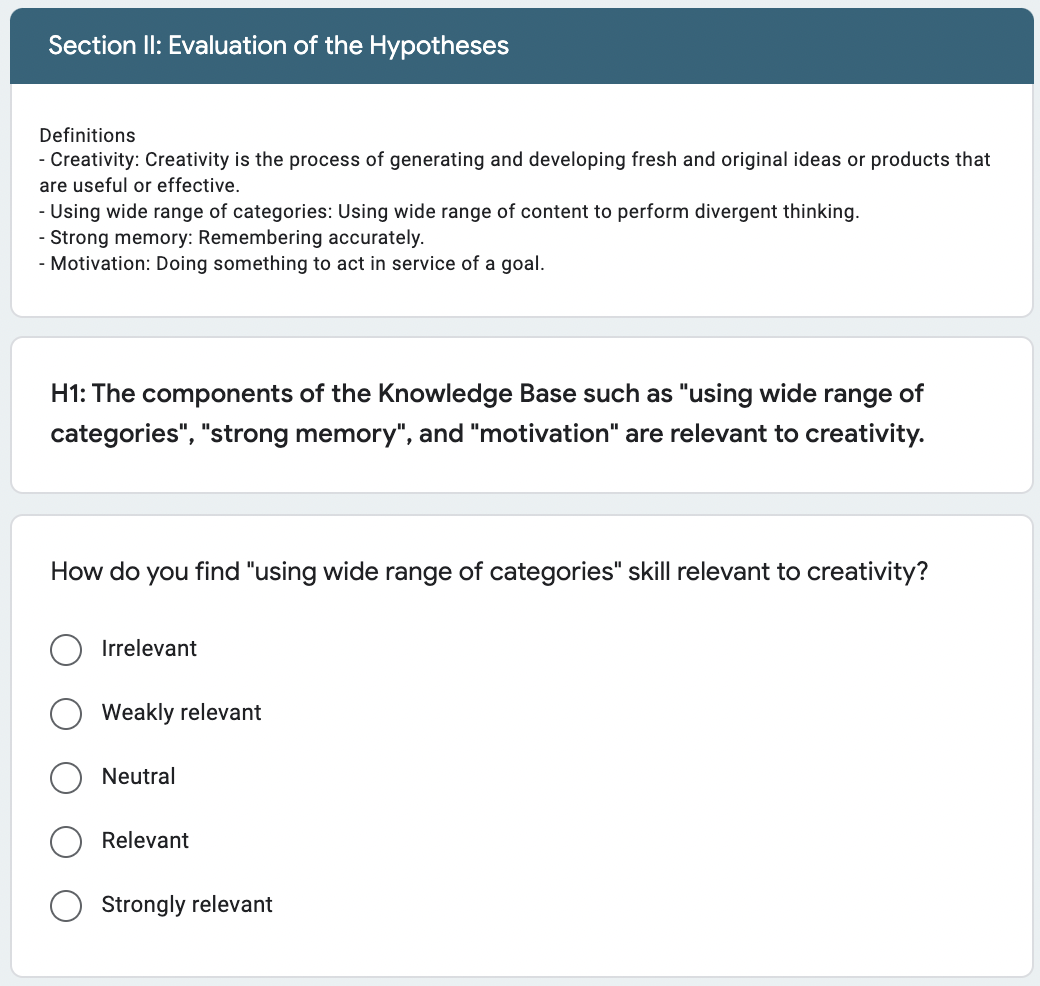} \\
    (a) & (b) \\
    \includegraphics[scale = 0.4]{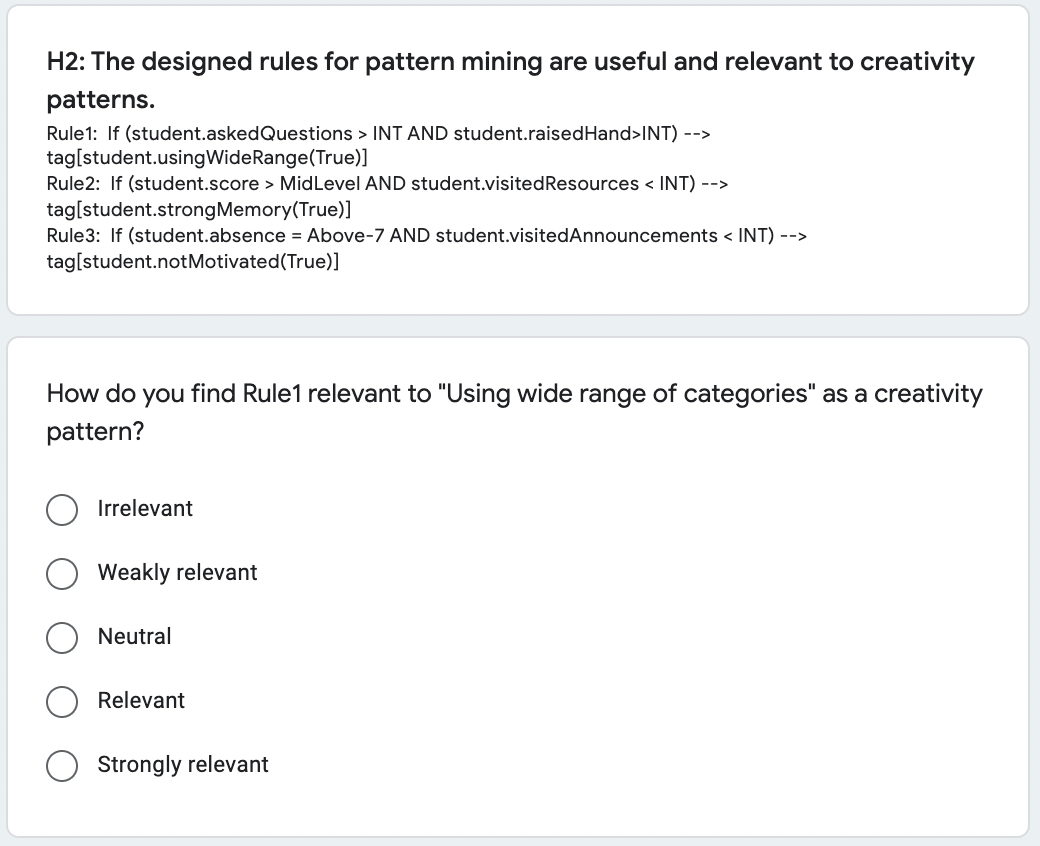} & \includegraphics[scale = 0.4]{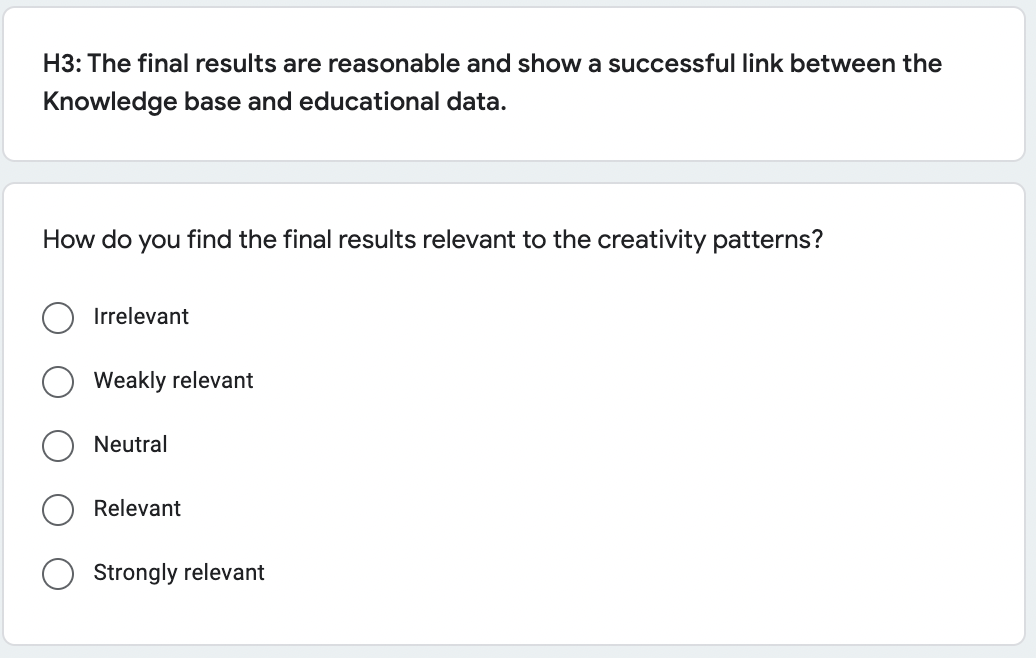} \\
    (c) & (d) \\
    \end{tabular}
    \caption{Examples of the questionnaire. (a) Demographic questions. (b) Evaluation of H1 for the "using wide range of categories" creativity pattern. (c) Evaluation of H2 for the "using wide range of categories" creativity pattern. (d) Evaluation of H3.}
    \label{Questions}
    \end{table}
\end{center}

\subsubsection{Experiment Results}
In this section, we aim to confirm or reject the hypotheses using the data acquired throughout the experiment.

\begin{figure}[t]
    \centering
    \includegraphics[width=1\textwidth]{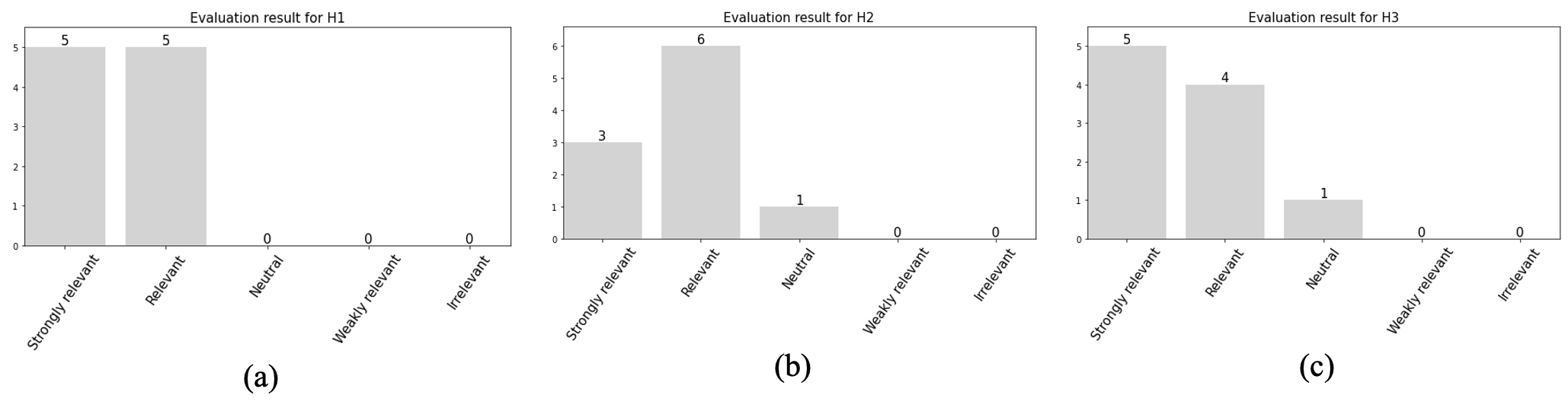}
    \caption{Evaluation of the hypotheses using the data acquired throughout the user study. (a) Evaluation of the concepts in the KB. (b) Evaluation of the defined rules. (c) Evaluation of the final results.}
    \label{H_evaluations}
\end{figure}

\begin{itemize}
    \item \textbf{Evaluation of H1}: H1 assumes that the components of the KB are relevant to creativity. The Figure \ref{H_evaluations}.a indicates that overall all the participants found that the linked concepts in the KB are relevant to creativity as a skill in education. Concepts such as Using a wide range of categories, Strong Memory, and Motivation have been investigated and rated by participants. All the participants found the concepts either strongly relevant or relevant to creativity.
    \item \textbf{Evaluation of H2}: H2 states that the designed rules for pattern mining are useful and relevant to creativity patterns. The Figure \ref{H_evaluations}.b  indicates that overall all the participants except one found the rules relevant to Using a wide range of categories, Strong Memory, and Motivation concepts in the taxonomy. Except for one, all the participants found the rules either strongly relevant or relevant to creativity.
    \item \textbf{Evaluation of H3}: H3 assumes that the query results are reasonable and show a successful link between the KB and educational data. The Figure \ref{H_evaluations}.c indicates that overall all the participants except one found that the model was successful in detecting those creativity patterns in the students. Except for one, all the participants found the results either strongly relevant or relevant to creativity.
\end{itemize}

\textbf{Discussion of Experiment Results.}
There are a few challenges towards validity, hence the findings of our study may not be regarded as definitive. They do, however, hint at a number of trends.
\begin{itemize}
    \item The finding of the user study supports the hypotheses H1, H2, and H3. However, H2, the rule-based pattern mining techniques require future improvement to gain higher score in the evaluation.
    \item The assigned timeframe for training the most of participants seems to be adequate except for two with no background in computing and education. 8 out of 10 participants successfully completed all four sections of the questionnaire in less than half an hour. Those participants with other backgrounds struggled to understand the related concepts and technical concepts. Hence, the training should be improved for future study cases.
    \item Based on our findings, mainly experts in education with knowledge, expertise, and interest in education and computing found our approach valid and confirmed the hypotheses.
\end{itemize}

Hence, based on the participants' feedback and the lessons learned, some future improvements could be considered for the approach. The motivation of participants, time pressure and training, and definition of the rules are important indicators that have a significant impact on the final evaluation results.
\begin{itemize}
    \item Motivation of Participants: It was completely voluntary to participate in the evaluation process which leads to acceptable engagement. However, in order to improve the evaluation, it is suggested to incentivize with nominal financial reward. As a consequence, participants will most likely be intrigued and passionately engaged in the process.
    \item Time Pressure and Training: The sessions for training and answering the questionnaire did not have a fixed time to minimize pressure. Still, some of the participants had a time conflict in their schedule and made them hurry towards the end. Although all managed to complete the tasks successfully, this may have impacted the way in which the tasks were completed and the questionnaire was answered. Hence, it is better to set a time limit for the tasks or separate the sessions to make it both flexible and more organized. This also can be helpful for those struggling participants with  a limited proficiency in education and computing and let them learn the concepts better and participate actively.
    \item Definition of the Rules: Our rule-based insight discovery approach, as the name implies, completely relies on the defined rules by educational experts. Thus, future improvements in terms of validity may include increasing the number of involved educational experts to define accurate rules. It is also beneficial to combine the rules with other existing rule mining techniques and compare the evaluation results with the outcomes of this study.
\end{itemize}
\section{Conclusion and Future Work}
This section highlights the paper's contributions and discusses possible future research areas.

\subsection{Conclusion}
Creativity is becoming a priority in the educational sector as it promotes cognitive complexity. Creativity requires extensive knowledge as well as the ability to use it effectively. Being creative entails experimenting with new possibilities in the pursuit of desired outcomes utilizing an existing set of knowledge or skills in a specific subject or setting. It takes time to develop and the process is more effective if students have certain knowledge and skills. The education system must provide both policy and educational methods to support students in gaining the required knowledge in their field of study. Teachers here can support creativity in their students by considering the ways that help students use their knowledge and come up with creative products. In the creative process, students require ongoing guidance and training. But before providing personalized training, it is needed to first detect creativity patterns in students.

In this paper, a data-driven technique is provided to relate students' behavior to creative thinking patterns and assist instructors to understand them from students' activities and assessment tasks. We concentrated on understanding the big educational data, used existing data curation techniques to turn the raw educational data into contextualized data and knowledge, built a domain-specific KB by leveraging the knowledge of education experts, and linked the contextualized data to the KB using a rule-based technique. As a result, we facilitated mining creative thinking patterns from contextualized data and knowledge and made it possible for educators to find students with creative thinking patterns. We also evaluated our approach through a user study, relying on the knowledge of education experts.

\subsection{Future Research Areas}
Regarding the information provided in this study, several strategies might be used to enhance the performance of the proposed method. This section lays out a plan to identify how these strategies may impact the system's performance in the future, as well as how they could be applied to the application of pattern mining within the education sector.

\subsubsection{Timely Analysis of Student Patterns}
Since features and attributes related to each student may change over time, it is beneficial to design a framework to enable educators monitor their students' performance. Age, level, grade, attendance, asking questions, discussion, score, and volunteering are examples of key attributes and features that will change over time, resulting in changes in previously observed patterns. This model may be expanded to include a dashboard considering all the important features over the time. By monitoring the features some alerts may be triggered and notify the educator to take action. Using this visualized system as a student profile, it will be possible for the educator to track their students' performance and behavior and plan for their education.

\subsubsection{Associate Rule Mining Technique}
Associate rule mining technique could be used to discover the relationship among educational features. It will uncover common patterns, relationships, or correlations among groupings of components or items in educational datasets. As future work, we intend to implement an associate rule mining technique to find meaningful relationships among important features in the process of identifying creativity patterns. In this way, we automate the process of creating rules and minimize the manual efforts needed to do so.

\subsubsection{Creativity Patterns}
In this study, we only investigated three patterns of creativity such as using a wide range of categories, strong memory, and motivation. We aim to continue this effort in the future and complete the identification of other key patterns. Self-esteem, social poise, and suspend judgment are other key patterns of creativity that all students should possess in education and are necessary to be investigated besides the current ones. Hence, to complete the study, we intend to expand the model and improve the data integration to finally be able to evaluate the students better.

\section*{Acknowledgements}
- I acknowledge the AI-enabled Processes (AIP\footnote{https://aip-research-center.github.io/}) Research Centre for funding My Master by Research project.

\bibliographystyle{abbrv}
\bibliography{references}

\end{document}